\pdfoutput=1
\documentclass[PAPER, atlasdraft=false, cernpreprint, english, texmf, orcidlogo]{atlasdoc}
 
\usepackage[subfigure]{atlaspackage}
\usepackage[backref=false]{atlasbiblatex}
\usepackage[BSM]{atlasphysics}

\graphicspath{{logos/}{figures/}}
 
\usepackage{ANA-SUSY-2019-24-PAPER-defs}

\AtlasTitle{A search for R-parity-violating supersymmetry in final states containing many jets in $pp$ collisions at \(\sqrt{s} = 13\,\text{TeV}\) with the ATLAS detector}

\AtlasAbstract{
A search for R-parity-violating  supersymmetry in final states with high jet multiplicity is presented.
The search uses $\SI{140}{\text{fb}^{-1}}$ of proton--proton collision data at $\sqrt{s} = \SI{13}{TeV}$ collected by the ATLAS experiment during Run 2 of the Large Hadron Collider.
The results are interpreted in the context of R-parity-violating supersymmetry models that feature prompt gluino-pair production decaying directly to three jets each or decaying to two jets and a neutralino which subsequently decays promptly to three jets.
No significant excess over the Standard Model expectation is observed and exclusion limits at the 95\% confidence level are extracted.
Gluinos with masses up to 1800 GeV are excluded when decaying directly to three jets. In the cascade scenario, gluinos with masses up to 2340 GeV are excluded for a neutralino with mass up to 1250 GeV.}

\AtlasRefCode{SUSY-2019-24}
 
\PreprintIdNumber{CERN-EP-2023-259}

\AtlasJournal{JHEP}
\AtlasJournalRef{JHEP 05 (2024 003)}
\AtlasDOI{DOI:10.1007/JHEP05(2024)003}

\hypersetup{pdftitle={ATLAS document},pdfauthor={The ATLAS Collaboration}}
 
\begin{document}
 
\maketitle

\section{Introduction}
% The next lines are included from the .//sections/Intro.tex input file
Supersymmetry (SUSY)~\cite{Golfand:1971iw,Volkov:1973ix,Wess:1974tw,Wess:1974jb,Ferrara:1974pu,Salam:1974ig}
is a theoretical extension of the Standard Model (SM) which proposes a fundamental symmetry that relates fermions and bosons
and predicts  a partner particle for each SM particle.
It is a promising theoretical possibility given its potential to solve the hierarchy problem~\cite{Sakai:1981gr,Dimopoulos:1981yj,Ibanez:1981yh,Dimopoulos:1981zb}.
An ad-hoc conserved quantity, R-parity~\cite{Farrar:1978xj}, is often introduced in SUSY models to avoid rapid proton decay,
rendering the lightest supersymmetric particle (LSP) stable and therefore a potential dark-matter candidate~\cite{Goldberg:1983nd,Ellis:1983ew}. However, there is no fundamental theoretical reason to impose strict R-parity conservation. R-parity-violating (RPV) SUSY models are well motivated and generally have fewer experimental constraints than many R-parity-conserving (RPC) models~\cite{Evans2014,Batell:2013zwa}. This suggests that the ATLAS Run 2 data sample could contain thousands of events where the supersymmetric partner of the gluon, the gluino ($\tilde{g}$) is present. This is plausible because gluino pairs are produced predominantly through the strong interaction, a process that can lead to a significantly large cross-section, depending on the gluino mass.
 
A search is presented for supersymmetric gluino pair production
with subsequent RPV decays into quarks in events with many jets using 140 $\text{fb}^{-1}$ of proton-proton ($pp$) collision data collected at $\sqrt{s}=13$ \TeV by the ATLAS detector during Run 2 of the Large Hadron Collider (LHC)\@.
Such a final state is predicted in RPV models with a non-zero baryon-number-violating \UDD coupling~\cite{DAmbrosio:2002ex,Csaki:2011ge}.
The dominant SM background process originates from multijet production, with a cross-section several orders of magnitude higher than the targeted signals.
Two approaches are implemented to distinguish between the SM background and potential SUSY signal.
The first, ``jet counting'' method, defines several signal regions, SR, requiring many high-$p_{\mathrm{T}}$ jets. The background is estimated from events containing low jet multiplicities and low momenta extrapolated to higher jet-momenta and multiplicities. To compensate the limitations of the simulated multi-jet background, data are used to normalize the simulation in several control regions.
The second, ``mass resonance'', approach aims to reconstruct the gluino mass with machine-learning methods, solving the combinatorial assignment challenge to correctly identify which jets belong to a given gluino. A mass-resonance search is then performed on the gluino-candidate mass spectrum. A fully data-driven approach is used to estimate the background, with a functional fit to the smoothly decreasing gluino-candidate mass distribution.
 
\begin{figure}[t]
\centering
\subfigure[]{\includegraphics[width=0.40\textwidth]{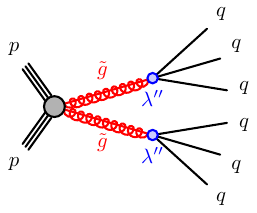}\label{fig:2by3}}
\subfigure[]{\includegraphics[width=0.40\textwidth]{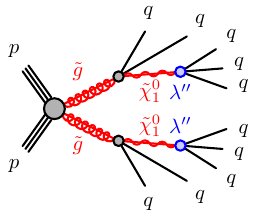}\label{fig:2by5}}
\caption{Signal diagrams for the \subref{fig:2by3} gluino direct decay and \subref{fig:2by5} cascade decay models.}
\label{fig:diagrams}
\end{figure}

Two RPV SUSY simplified signal models~\cite{Alwall:2008ve,Alwall:2008ag,Alves:2011wf} featuring gluino-pair production are targeted. Figure~\ref{fig:2by3} presents the gluino direct decay model, where the gluino decays into three quarks via a $\lambda^{''}_{ijk}$ \UDD RPV coupling, leading to final states containing at least six jets. The indices $i, j, k $ denote the generations of the quarks involved in the interaction. The gluino cascade decay model  is presented in Figure~\ref{fig:2by5}, in which the gluino decays into two quarks and neutralino, $\tilde{\chi}_{1}^{0}$, where the neutralinos result from the mixing between the supersymmetric partners of the neutral SM bosons. The neutralino then decays into three quarks, again via the $\lambda^{''}_{ijk}$ \UDD coupling, leading to at least ten jets in the final state. Both the scenarios assume that $\lambda^{''}_{ijk}$ is large enough to ensure prompt SUSY decays. Two couplings are considered, $\lambda^{\prime\prime}_{112}$ and $\lambda^{\prime\prime}_{113}$, leading to the RPV decays $\tilde{g}/\tilde{\chi_{1}^{0}}\rightarrow uds$ referred to as the UDS-decay or $\tilde{g}/\tilde{\chi_{1}^{0}}\rightarrow udb$ referred to as the UDB-decay respectively. The results of this article apply equally to other couplings, $\lambda_{ij2}''$, $\lambda_{ij3}''$, with $i,j \in {2,3}$, since it has the same experimental final state. The UDB-decay has a unique signal phenomenology containing bottom quarks and a dedicated event selection is employed to specifically target this scenario.

Previous searches in this final state were performed by the ATLAS~\cite{SUSY-2013-07,SUSY-2016-22} and CMS collaborations~\cite{CMS-EXO-17-030}.
New methods  are used here for both the jet counting and mass resonance approaches, dramatically improving the sensitivity beyond the expected gains due to the larger data sample.

% End of text imported from the .//sections/Intro.tex input file

\section{ATLAS detector}
% The next lines are included from the .//sections/Detector.tex input file
\newcommand{\AtlasCoordFootnote}{
ATLAS uses a right-handed coordinate system with its origin at the nominal interaction point (IP)
in the centre of the detector and the \(z\)-axis along the beam pipe.
The \(x\)-axis points from the IP to the centre of the LHC ring,
and the \(y\)-axis points upwards.
Polar coordinates \((r,\phi)\) are used in the transverse plane,
\(\phi\) being the azimuthal angle around the \(z\)-axis.
The pseudorapidity is defined in terms of the polar angle \(\theta\) as \(\eta = -\ln \tan(\theta/2)\).
Angular distance is measured in units of \(\Delta R \equiv \sqrt{(\Delta\eta)^{2} + (\Delta\phi)^{2}}\).}

The ATLAS experiment~\cite{PERF-2007-01} at the LHC is a multipurpose particle detector
with a forward--backward symmetric cylindrical geometry and a near \(4\pi\) coverage in
solid angle.\footnote{\AtlasCoordFootnote}
It consists of an inner tracking detector surrounded by a thin superconducting solenoid
providing a \qty{2}{\tesla} axial magnetic field, electromagnetic and hadron calorimeters, and a muon spectrometer.
The inner tracking detector covers the pseudorapidity range of \(|\eta| < 2.5\).
It consists of silicon pixel, silicon microstrip, and transition radiation tracking detectors.
Lead/liquid-argon (LAr) sampling calorimeters provide electromagnetic (EM) energy measurements
with high granularity.
A steel/scintillator-tile hadron calorimeter covers the central pseudorapidity range of (\(|\eta| < 1.7\)).
The endcap and forward regions are instrumented with LAr calorimeters
for both the EM and hadronic energy measurements up to \(|\eta| = 4.9\).
The muon spectrometer surrounds the calorimeters and is based on
three large superconducting air-core toroidal magnets with eight coils each.
The field integral of the toroids ranges between \num{2.0} and \qty{6.0}{\tesla\metre}
across most of the detector.
The muon spectrometer includes a system of precision tracking chambers and fast detectors for triggering.
A two-level trigger system is used to select events.
The first-level trigger is implemented in hardware and uses a subset of the detector information
to accept events at a rate of less than \qty{100}{\kHz}.
This is followed by a software-based trigger that
reduces the accepted event rate to \qty{1}{\kHz} on average
depending on the data-taking conditions.
An extensive software suite~\cite{ATL-SOFT-PUB-2021-001} is used in data simulation, in the reconstruction
and analysis of real and simulated data, in detector operations, and in the trigger and data acquisition
systems of the experiment.

% End of text imported from the .//sections/Detector.tex input file

\section{Data and simulated event samples}
% The next lines are included from the .//sections/Data_MC.tex input file
The data analysed were collected between 2015 and 2018 at a centre-of-mass energy of 13~\TeV\ with a 25~ns proton bunch crossing interval.
The average number of $pp$ interactions per bunch crossing, referred to as pile-up, ranged from 13 in 2015 to around 38 in 2017--2018.
After applying conditions on the beam, detector and data-quality~\cite{Aad:2700249} the data sample has a total integrated luminosity of 140~fb$^{-1}$.
 
Monte Carlo (MC) samples are primarily used in the analysis to estimate the expected number of events for a given signal scenario. MC samples are also used to aid in the modelling of the SM backgrounds in the signal regions, or as a cross-check of the data-driven methods used to model the expected background yield.
 
Signal samples were simulated at leading-order (LO) accuracy with up to two additional partons using the \MGNLO event generator~\cite{Alwall:2011uj} interfaced with \PYTHIA[8.235]~\cite{Sjostrand:2014zea}. The A14~\cite{ATL-PHYS-PUB-2014-021} set of tuned parameters was used for the underlying event together with the \NNPDF[2.3lo]~\cite{Ball:2012cx} parton distribution function (PDF) set. The \EVTGEN program~\cite{Lange:2001uf} was used to model the decays of heavy-flavour hadrons. The generated events are then passed through a simulation~\cite{SOFT-2010-01} of the ATLAS detector geometry and response using \GEANT~\cite{Geant4PhysicsLists}. The signal cross-sections were calculated at next-to-next-to-leading-order (NNLO) in the strong coupling constant, adding the resummed soft gluon emission at next-to-next-to-leading-logarithm accuracy (NNLO+NNLL)~\cite{Beenakker:2016lwe,Beenakker:2014sma,Beenakker:2013mva,Beenakker:2011sf,Beenakker:2009ha,Kulesza:2009kq,Kulesza:2008jb,Beenakker:1996ch}.

Multijet events constitute the dominant background in the search region. Multijet production in the SM was simulated using \PYTHIA[8.230]~\cite{Sjostrand:2014zea} with LO matrix elements for dijet production which are matched to the parton shower.
The renormalisation and factorisation scales were set to the geometric mean of the squared transverse masses of the two outgoing particles in the matrix element. The \NNPDF[2.3lo] PDF set was used in the matrix element generation, the parton shower, and the simulation of the multi-parton interactions. The A14 set of tuned parameters is used. Perturbative uncertainties were estimated through event weights~\cite{Mrenna:2016sih} that encompass variations of the scales at which the strong coupling constant is evaluated in the initial- and final-state shower and the PDF uncertainty in the shower and the non-singular part of the splitting functions. A similar method was used to estimate the uncertainties for the signal modelling.

In the regions requiring the presence of a jet containing $b$-hadrons ($b$-tagged), there is a contribution from top-quark pair production ($t\bar{t}$). The production of fully hadronic decays of $t\bar{t}$ events, was modelled at NLO using the \POWHEGBOX~\cite{Alioli:2010xd} generator. Additional $t\bar{t}$ samples were simulated with  \MGNLO interfaced with \PYTHIA[8], and with \POWHEGBOX interfaced with \HERWIG[7]~\cite{Bahr:2008pv,Bellm:2015jjp}, for the evaluation of systematic uncertainties.
 
The effect of pile-up interactions was modelled by overlaying the simulated hard-scattering event with inelastic \(pp\) events simulated with \PYTHIA[8.186]~\cite{Sjostrand:2007gs} using the \NNPDF[2.3lo] set of parton distribution functions (PDF) and the A3 set of tuned parameters~\cite{ATL-PHYS-PUB-2016-017}. The simulated events are weighted to reproduce the distribution of the average number of interactions per bunch crossing (\(\left<\mu \right>\)) observed in the data. The \(\left<\mu \right>\) value in data is rescaled by a factor of \(1.03\pm 0.04\) to improve agreement between data and simulation in the visible inelastic \(pp\) cross-section~\cite{STDM-2015-05}.

% End of text imported from the .//sections/Data_MC.tex input file

\section{Event reconstruction}
% The next lines are included from the .//sections/Objects.tex input file
As the signal scenarios under investigation have a general event phenomenology with many energetic jets, an identical trigger strategy and a set of common object definitions can be used for both the analysis strategies.
 
Events are required to satisfy an $H_{\mathrm{T}}$ trigger, where $H_{\mathrm{T}}$ is defined as the scalar sum of the transverse momentum ($p_{\mathrm{T}}$) of the jets, identified on the trigger level, in the event.
To ensure the trigger is fully efficient, a selection of $H_{\mathrm{T}} > 1100$~\GeV and a selection on the $p_{\mathrm{T}}$ of the leading jet, $p_{\mathrm{T}}(j_1) > 200$\,\GeV, and the 4$^{\text{th}}$ leading jet, $p_{\mathrm{T}}(j_4) > 50$~\GeV, are applied.
The $H_{\mathrm{T}}$ trigger has an efficiency exceeding 95\% for signal events satisfying the selection for all the data-taking periods.

The constituents for the analysis-level jet reconstruction are identified by combining measurements from both the inner-detector system (ID) and calorimeter using a  particle flow (PFlow) algorithm~\cite{PERF-2015-09}, which suppresses calorimeter energy deposits arising from charged pile-up particles and takes the momentum estimate from tracks whenever the tracker resolution is better than the calorimeter resolution. These jets are defined
using the anti-k$_{\mathrm{T}}$ algorithm~\cite{Cacciari:2008gp, Fastjet} with a size parameter of $R=0.4$. They are calibrated using simulation with corrections obtained by using in situ techniques in data~\cite{JETM-2018-05}.
Jets containing a large particle momentum contribution from pile-up vertices, as measured by the jet vertex tagger (JVT) discriminant~\cite{PERF-2014-03} are rejected if they have $p_{\mathrm{T}} \in [20,60]$\,\GeV, $|\eta| < 2.4$ and a discriminant value of JVT $<$ 0.5.
Two classes of jets are defined: ``baseline'' jets and ``signal'' jets. Baseline jets require $p_{\mathrm{T}} > 20$~\GeV and $|\eta| < 4.8$. Signal jets are used for the computation of kinematic variables and for the final event selections and require $p_{\mathrm{T}} > 50$~\GeV and $|\eta| < 2.8$.
 
Selected jets are tagged as $b$-jets if they are within the inner tracking detector acceptance of $|\eta| =  2.5$ and are identified by a
multivariate algorithm (DL1r) which uses a selection of inputs including information about the impact parameter
of tracks, the presence of displaced secondary vertices and the reconstructed flight paths of
$b$- and $c$-hadrons inside the jet~\cite{FTAG-2019-07}. The $b$-tagging algorithm uses a working point with an efficiency of
77\% to tag a $b$-quark jet, determined with a sample of simulated $t\bar{t}$ events. The corresponding misidentification (mis-tag) rate
is 20\% for $c$-jets and 0.9\% for light-flavour jets. Differences between data
and MC simulation for the efficiency and mis-tag rate are taken into account with
correction factors as described in Ref.~\cite{PERF-2013-06}.
 
As the signal scenarios considered do not contain any light leptons ($e$, $\mu$) signal sensitivity can be increased by rejecting events containing leptons. Electron candidates are reconstructed from an isolated electromagnetic calorimeter energy deposit matched to an inner detector track~\cite{EGAM-2018-01} and are required to have $p_{\mathrm{T}} > 10$~\GeV and $|\eta| < 2.47$, and to satisfy the ``Loose'' likelihood-based identification criteria described in Ref.~\cite{EGAM-2018-01}. Muon candidates are formed by combining information from the muon spectrometer and inner detector as described in Ref.~\cite{PERF-2015-10} and are required to have $p_\mathrm{T} > 10$~\GeV and $|\eta| < 2.7$. Furthermore, muon candidates must satisfy the ``Medium'' identification requirements described in Ref.~\cite{PERF-2015-10}. In both the cases lepton candidates must additionally have a longitudinal impact parameter relative to the primary vertex with $|z_0 \sin{\theta}| < 0.5$~mm.
 
Scale factors are applied to simulated events to account for differences between data and simulation for reconstruction, identification and isolation efficiencies. Similar corrections are also applied to the probability of mis-tagging jets originating from the hard scattering as pile-up jets with the JVT discriminant, and the  corrections related to the efficiency of identifying jets arising from $b$-hadrons.
 
After the object selection step, a procedure to avoid double counting of tracks and energy deposits matched to overlapping reconstructed jets,  electrons, and muons is implemented. This procedure applies the following actions to the baseline jets and leptons in a sequential order. If an electron lies within $\Delta R  = \sqrt{(\Delta\eta)^{2} + (\Delta\phi)^{2}}= 0.2$ of the jet axis, then the jet is removed, whereas if jet and an electron are within $\Delta R = 0.4$, then the electron is removed. If a jet and a muon are within $\Delta R = 0.2$, or the muon track is matched to the jet, then the jet is considered for removal. If the number of tracks within the jet is fewer than three and if the jet and total track $p_{\mathrm{T}}$ is consistent with the muon energy, then the jet is removed. Finally, if a muon and a jet are within $\Delta R = 0.4$ then the muon is removed.

% End of text imported from the .//sections/Objects.tex input file

\label{sec:Event_Reco}
 
\section{Analysis strategy}
% The next lines are included from the .//sections/Analysis.tex input file
\begin{figure}[t]
\centering
 
\subfigure[]{\includegraphics[width=0.49\textwidth,trim={0cm 0cm 0 0},clip]{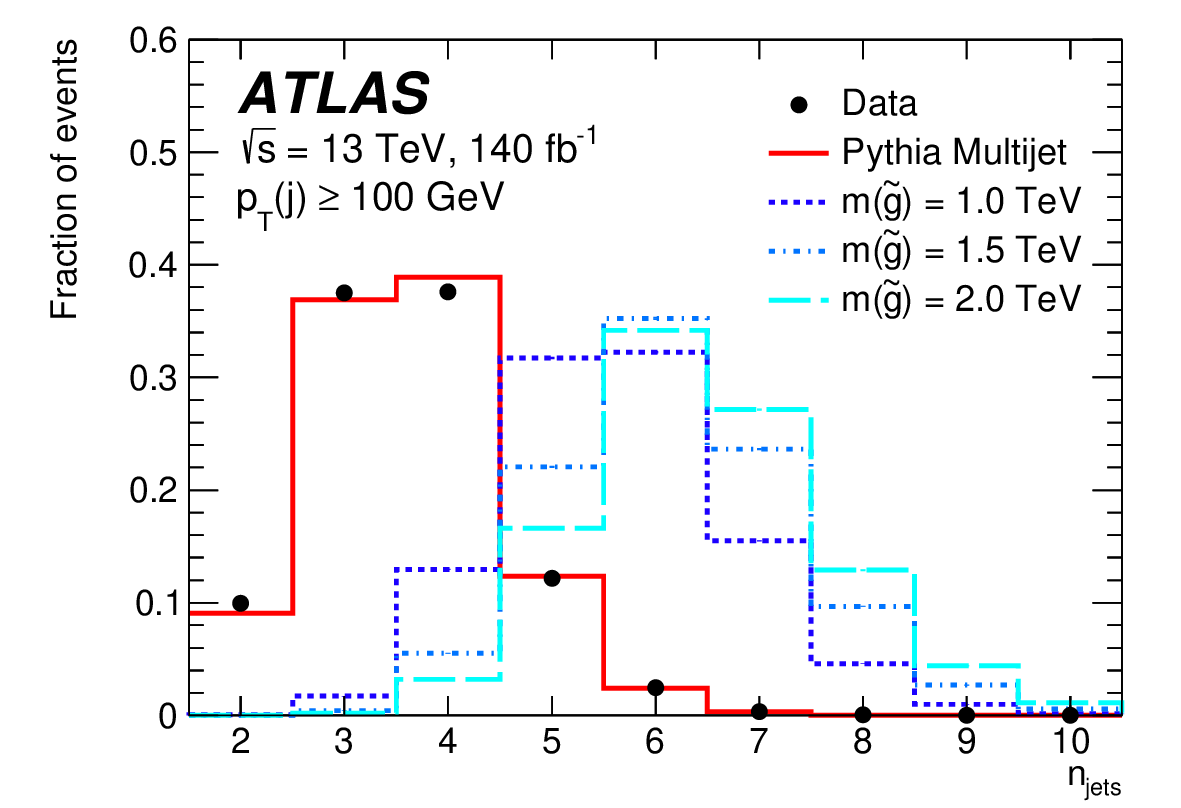}\label{fig:var-shapes_nj}}
\subfigure[]{\includegraphics[width=0.49\textwidth,trim={0cm 0cm 0 0},clip]{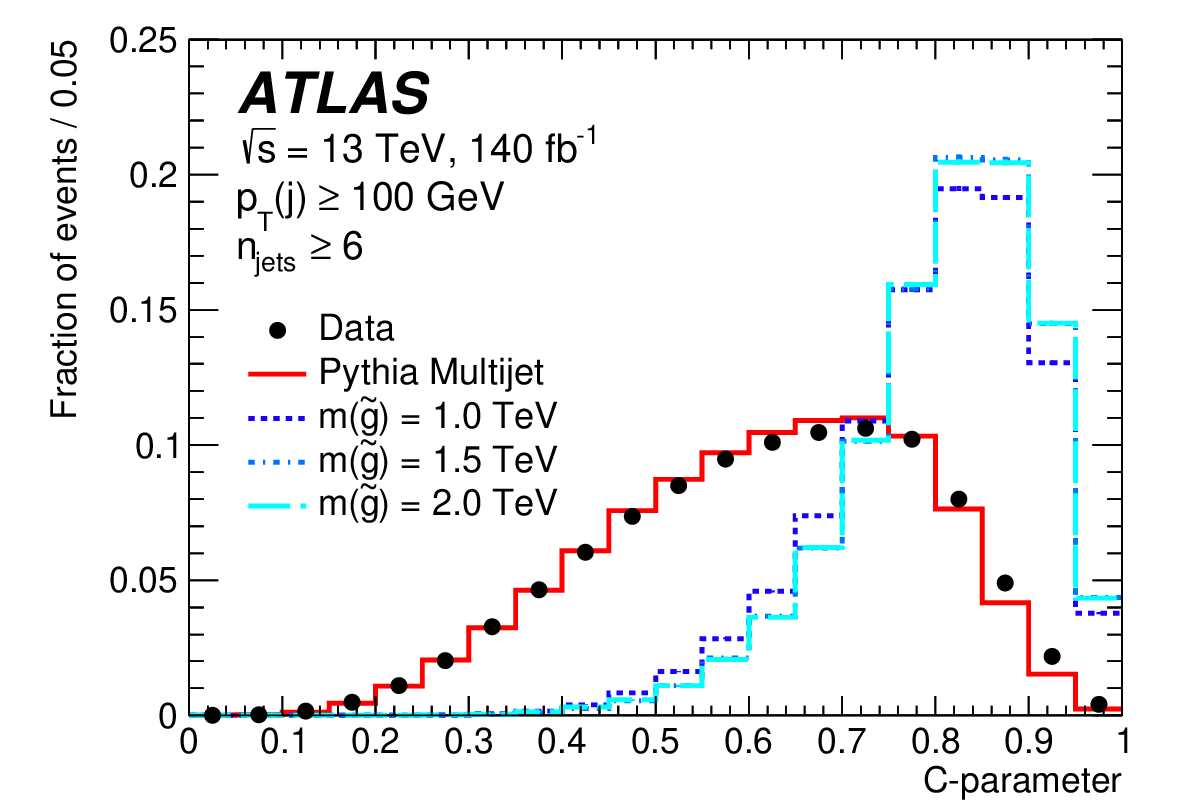}\label{fig:var-shapes_c}}
 
\caption{
Comparison between distributions, normalised to unity,  of the observed data, the QCD multijet background and signal models.
Figure \subref{fig:var-shapes_nj}: $n_{\textrm{jets}}$ spectrum with $p_{\mathrm{T}} \geq 100$~\GeV.
Figure \subref{fig:var-shapes_c}: distribution of the $C$ variable for events with at least six jets with $p_{\mathrm{T}} \geq 100\,$~\GeV.
}
\label{fig:var-shapes}
\end{figure}
 
Two complementary analysis methods are used for the signal scenarios.
In addition to possessing common object and trigger requirements, similar kinematic variables are employed to do a simple discrimination between signal and background, which the two methods build upon.
 
Events are required to satisfy the $H_{\mathrm{T}}$ trigger, and all of the requirements needed to assure the trigger is fully efficient as discussed in Section~\ref{sec:Event_Reco}.
All accepted events are required to have no leptons ($e$, $\mu$), and are required to contain at least four jets with $p_{\mathrm{T}} > 50$~\GeV.
 
For both the methods, the event-shape variable $C$~\cite{ATLAS:2020vup} derived from the linearized sphericity tensor of the event is used to distinguish between signal and background.
The sphericity tensor, which captures the momentum distributions in an event, can be reduced to three eigenvalues, $\lambda_1,\lambda_2$, and $\lambda_3$, representing the shape of this distribution along three orthogonal directions.
The $C$ variable is calculated as a combination of these eigenvalues:
\begin{equation}
C = 3(\lambda_1\lambda_2+\lambda_1\lambda_3+\lambda_2\lambda_3) \,.
\label{eq:C_value}
\end{equation}
For events consisting of two back-to-back jets, which dominate the multijet background from quantum chromodynamic interactions, QCD, the $C$ value tends to be smaller than for gluino decays where the energy is distributed more uniformly or isotropically and the $C$ value tends to larger values.
A selection on the $C$ variable and a selection requiring high multiplicity of jets with $p_{\mathrm{T}}$ above a given threshold ($n_{\text{jets}}$) are employed by both the analysis methods as key variables to discriminate between the signal and the SM background.
Figure \ref{fig:var-shapes} presents unit-area-normalised comparisons of these discriminating variables, showing the significant differences between the signal models and the background.
 
The following two sections describe the signal optimisation strategy and background estimation techniques
for the \emph{jet counting} and \emph{mass resonance method} respectively.
While both the methods target the direct decay scenario, the jet counting method provides a more model-independent approach
which could detect a general excess in events with large jet multiplicities and also investigates the cascade decay scenario;
the mass resonance method is more model specific, and focuses exclusively on the direct decay scenario,
seeking to reconstruct the gluino mass directly from the decay products.
 
\subsection{Jet counting method}

% The next lines are included from the .//sections/Analysis_JC.tex input file
The jet counting method is built on the fact that the signal scenarios considered produce a large multiplicity of high-$p_{T}$ jets, a feature that was already exploited in previous analyses targeting similar models~\cite{SUSY-2012-16,SUSY-2013-07}.
In this approach, signal regions are defined by requiring a high jet multiplicity and a tight requirement on the $p_\mathrm{T}$ of the jets. The expected number of background events in this region is estimated by using control regions (CRs) which are defined with lower jet $p_{\mathrm{T}}$ requirements, which are then extrapolated to the SRs. To check the validity of this extrapolation, the background expectation (extrapolated from the CR) is compared with the observed data in intermediate validation regions (VRs), which are tighter than the CR requirements, but looser than the SRs.
 
In total seven SRs are defined to target different regions of the SUSY parameter space under consideration.
The SRs are sensitive to both the direct gluino decay scenario and the cascade scenario.
Table \ref{tab:cca-SR} presents the SR selections.
All SRs require at least seven high-$p_{\mathrm{T}}$ jets, and tight selections on the $C$ variable.
A selection of at least seven jets is chosen, instead of six as suggested by the signal diagram from Figure~\ref{fig:2by3},
as the inclusion an extra jet which arises from initial- or final-state radiation is found to increase the sensitivity to the signal scenarios and further reject background events.
Two SRs are defined with a requirement on the number of $b$-tagged jets present to specifically target the scenarios where the UDB-coupling allows  $b$-quarks in the decay.

\begin{table}[ht]
\caption{\label{tab:cca-SR}
SR definitions for the jet counting method, where $n_{\mathrm{jets}}$ represents the number of jets above the given \pT{} threshold ($p_{\mathrm{T}}(j)$).
The common analysis selections on the $H_{\mathrm{T}}$, $p_{\mathrm{T}}(j_1)$ and trigger selection are also applied. }
\centering{
\begin{tabular}{l @{\hspace{3em}} C{3em} C{3em} C{3em} C{3em}}
\toprule
& $n_{\mathrm{jets}}$ & $p_{\mathrm{T}}(j)$ [GeV] & $C$ & $n_{b-\mathrm{jets}}$ \\[5pt]
\midrule[0.2pt]
SR1 & $\geq 7$ & $\geq 180$ & $\geq 0.90$ & -- \\[5pt]
SR2 & $\geq 7$ & $\geq 220$ & $\geq 0.90$ & -- \\[5pt]
SR3 & $\geq 7$ & $\geq 240$ & $\geq 0.90$ & -- \\[5pt]
SR4 & $\geq 8$ & $\geq 180$ & $\geq 0.85$ & -- \\[5pt]
SR5 & $\geq 8$ & $\geq 210$ & $\geq 0.85$ & -- \\[5pt]
SR1bj & $\geq 7$ & $\geq180$ & $\geq 0.85$ & $\geq 2$ \\[5pt]
SR2bj & $\geq 8$ & $\geq180$ & $\geq 0.85$ & $\geq 2$ \\
\bottomrule
\end{tabular}
}
\end{table}
 
\subsubsection*{Background estimation}
The primary source of background arises from QCD multijet events.
In large jet multiplicities, the MC simulation alone cannot provide accurate descriptions of the event kinematics.
The $2\rightarrow 2$ LO matrix element calculations and the parton shower method are limited in the modelling of the absolute rate as a function of the number of jets.
However, thanks to the tunings performed~\cite{ATL-PHYS-PUB-2019-017},
the shape of the $n_\text{jets}$ distribution and the event shape variables
in data are reasonably well described~\cite{ATLAS:2020vup,STDM-2020-20}.
Also, a minimal set of analysis selection variables is used, as shown in Table~\ref{tab:cca-SR}, reducing the reliance on the MC.
To further address simulation shortcomings, we employ a semi-data-driven approach,
following a background determination method similar to previous searches~\cite{SUSY-2012-16,SUSY-2013-07}.
The expected number of  events with a given jet multiplicity $n_\text{jets}=i$
above a certain jet-$p_{\mathrm{T}}$ threshold $X$ (denoted by $N_{i, \pt^X}$) can be evaluated as:
\begin{equation}
\begin{split}
N_{i,\pT^X} = w_{i} \cdot N_{4, \pT^X}^{\text{Data}} \cdot \frac{N_{i, \pT^X}^{\text{MC}}}{N_{4, \pT^X}^{\text{MC}}} \,, \\
w_{i}= {\frac{N^{\text{Data}}_{i,\pT^{60}}}{N^{\text{Data}}_{4,\pT^{60}}}}\bigg/{\frac{N^{\text{MC}}_{i,\pT^{60}}}{N^{\text{MC}}_{4,\pT^{60}}}} \,.
\end{split}
\label{eq:cca-bkg_eg}
\end{equation}
 
Multijet MC is employed to compute transfer factors across different jet multiplicities $\bigl( N_{i, \pT^X}^{\text{MC}} / N_{4, \pT^X}^{\text{MC}}\bigr)$.
The prediction is normalised using a signal-free data region $\bigl( N_{4, \pT^X}^{\text{Data}} \bigr)$.
To address potential biases in the MC modelling of the jet multiplicity, correction factors ($w_{i}$) are obtained from CRs, based on the ratio of transfer factors computed separately in data and MC.
In these CRs, the jet $p_{\mathrm{T}}$ threshold is reduced to 60~\GeV, to avoid eventual signal contamination.
However, the same selection on the variable $C$ as in the corresponding SR is used.
The SRs are inclusive in jet multiplicity ($\geq n_{\mathrm{jets}}$);
event yields are then estimated by summing exclusive jet multiplicities up to $n+2$,
as successive terms of the series have negligible contributions ($\leq 0.1\%$):
 
\begin{equation}
N_{\geq n,\pT^X} = \sum_{i=n}^{n+2} N_{i,\pT^X} \,.
\end{equation}

\begin{figure}[!ht]
\centering
\subfigure[]{\includegraphics[width=0.49\textwidth]{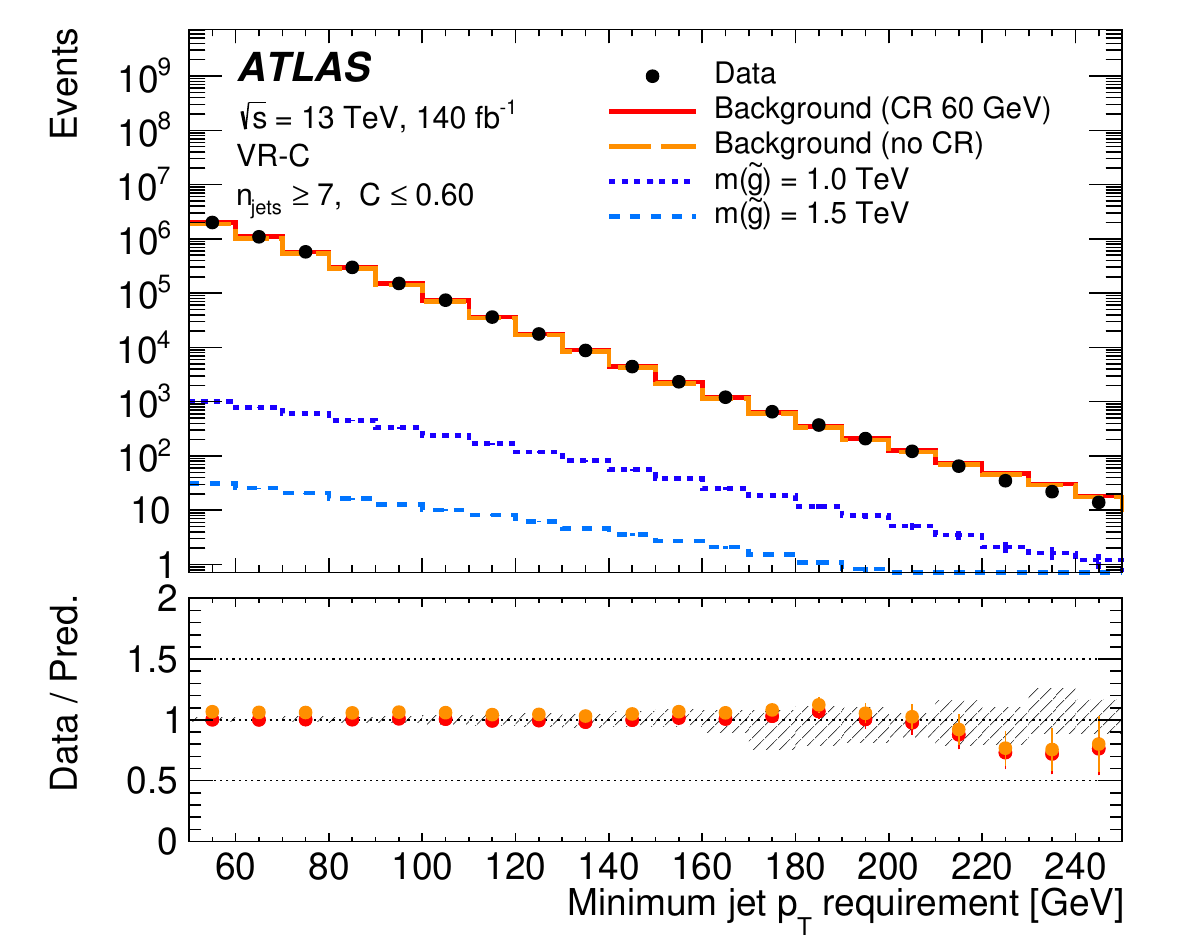} \label{fig:jc-bkg-val_a}}
\subfigure[]{\includegraphics[width=0.49\textwidth]{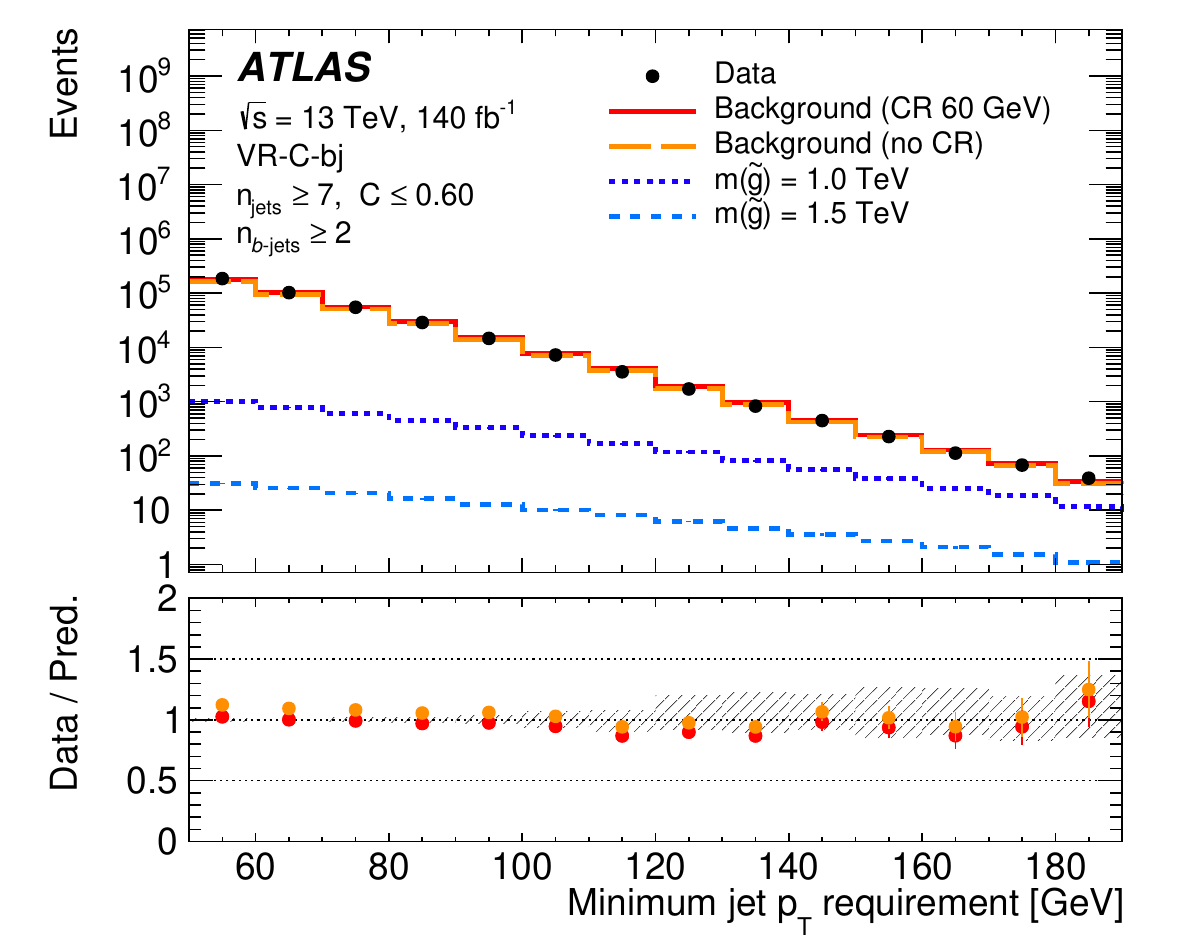} \label{fig:jc-bkg-val_b}}
 
\caption{
Test of the background estimation for different minimum jet $p_{\mathrm{T}}$ requirements \subref{fig:jc-bkg-val_a} without and \subref{fig:jc-bkg-val_b} with $b$-tagging selections.
The solid red line is the prediction using correction factors estimated in the CR, while the orange line is obtained without such corrections.
The bottom panels show the ratio of the data to the predicted yields, with (red dots) and without (orange dots) the CR\@.
The hatched pattern in the bottom panels represents the total systematic uncertainty in the background estimate when the CR is used.
}
\label{fig:jc-bkg-val}
\end{figure}

When introducing the selection on the number of $b$-tagged jets ($n_{b-\mathrm{jets}}$), as in SR1bj and SR2bj, there is a sizeable contribution from the \ttbar SM process, with up to 30\% of the total background consisting of \ttbar.
In this case, the $N^{\text{MC}}$ terms in Eq.~(\ref{eq:cca-bkg_eg}) are treated as the combined sum of multijet and $t\bar{t}$ events.
Figure \ref{fig:jc-bkg-val} shows two examples of the background estimation method with and without the selection on $b$-tagged jets.
The latter figure also shows the small impact of the correction factors $w_i$.
 
Signal injection tests were conducted to study the effect of signal contamination in the CRs.
The latter would bias the background estimate but have a minor impact on the analysis sensitivity.
The amount of bias is inversely proportional to the gluino mass.
However, it is always smaller than the other systematic uncertainties associated with the method
and vanishes rapidly to less than 1\% for models with masses above $1.1$~\TeV.

\subsubsection*{Background validation}
 
\begin{figure}[!ht]
\centering
\subfigure[]{\includegraphics[width=0.49\textwidth]{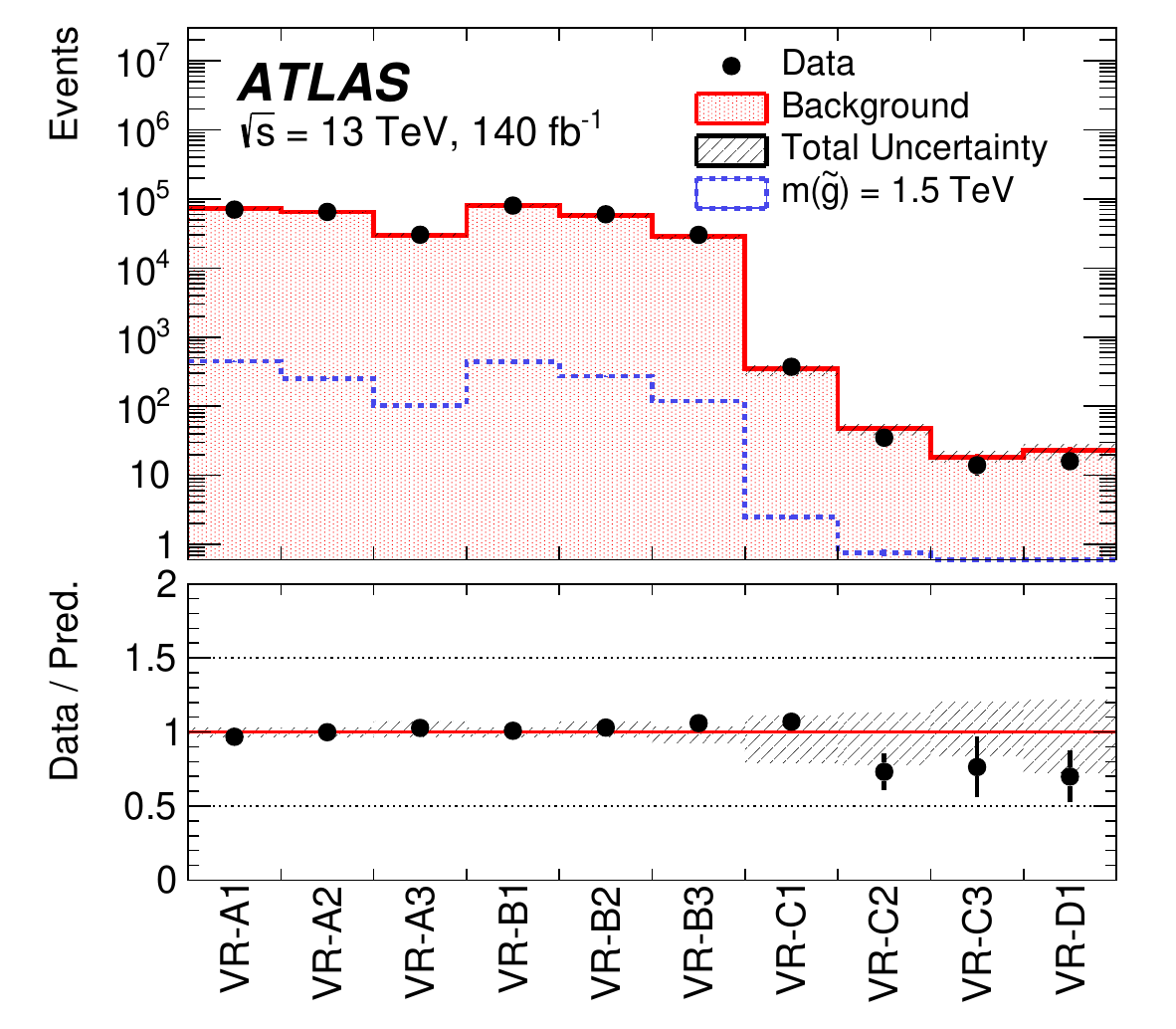}\label{fig:cca-closure_a}}
\subfigure[]{\includegraphics[width=0.49\textwidth]{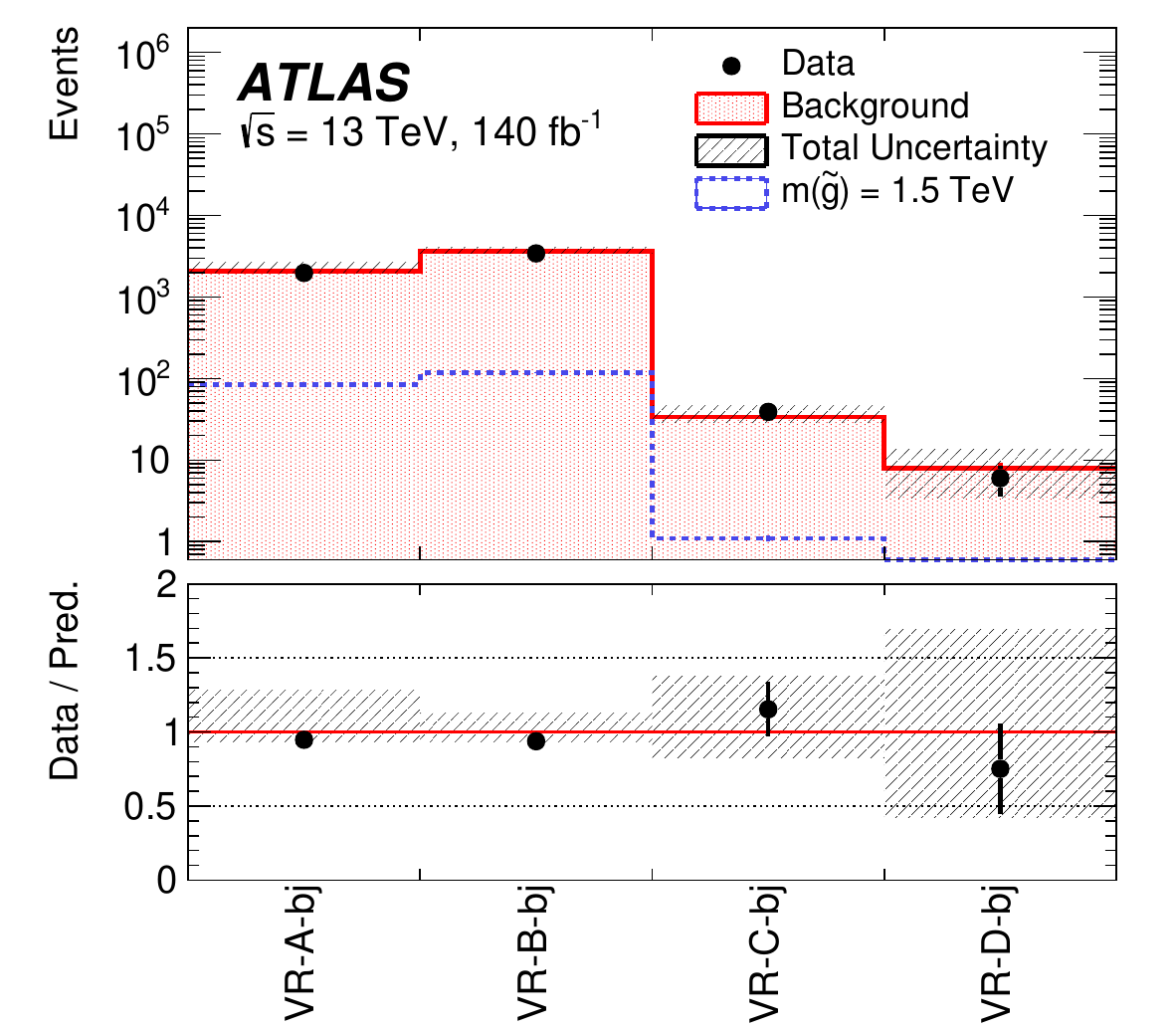}\label{fig:cca-closure_b}}
 
\caption{
Comparison of event yields between the observed data and the background expectation in the VRs.
Figure~\subref{fig:cca-closure_a}: VRs containing no explicit requirement on the number of $b$-tagged jets.
Figure~\subref{fig:cca-closure_b}: VRs containing at least two $b$-tagged jets.
The bottom panel presents the ratio of data to the background prediction.
The hatched pattern represents the combined statistical and systematic uncertainty in the background estimate.
}
\label{fig:cca-closure}
\end{figure}

To evaluate the background modelling several VR sets are defined.
They are presented in Tables \ref{tab:cca-VR} and \ref{tab:cca-VR-bj-final}.
The VRAs and VRBs are designed to validate the method at high values of $C$, but with lower jet multiplicities to negate signal contamination.
The VRCs and VRDs validate the method in a high jet multiplicity and jet momenta region, utilising an inverted $C$ requirement compared with minimal signal contamination.
The individual regions in a given VR set are not orthogonal, as the regions only differ in the \pT{} selection and can therefore be considered to be subsets of each other.
Four dedicated VRs (VR-A-bj, VR-B-bj, VR-C-bj and VR-D-bj) are defined to validate the background modelling for the selections requiring at least two $b$-tagged jets,
due to the different background composition when $b$-jets are present in the final state.
Figure \ref{fig:jc-bkg-val} shows the agreement between the background estimate in the VR-C and VR-C-bj as a function of the jet $p_{\mathrm{T}}$ threshold used to count the jets.
It is seen that there is generally acceptable agreement between the data and the background estimate while increasing the threshold to the highest values used in the SRs.
Figure \ref{fig:cca-closure} presents the yields of the VRs, displaying the agreement between the background expectation from the jet counting method and the observed data.
A slight discrepancy is observed in VR-B3; this inconsistency is used to define an additional non-closure uncertainty of 5\% on the expected background yields in the SRs.
 
\begin{table}[ht]
\caption{\label{tab:cca-VR}
VR definitions and yields, for the regions used to validate the background strategy without an explicit selection on the number of $b$-tagged jets. The common analysis selections on the $H_{\mathrm{T}}$, $p_{\mathrm{T}}(j_1)$ and trigger selection are also applied. The uncertainties shown contain both the statistical and systematic uncertainties.
}
\centering{
\begin{tabular}{c @{\hspace{3em}} C{3em} C{3em} C{4em}  @{\hspace{2.5em}} R{4.8em} @{\hspace{\arrayrulewidth}} L{3.2em}  r}
\toprule
& $n_{\mathrm{jets}}$ & $p_{\mathrm{T}}(j)$ [GeV] & $C$ & \multicolumn{2}{C{8em}}{Background Expectation} & \multicolumn{1}{C{3.5em}}{$\quad$Data} \\
\midrule[0.2pt]
VR-A1 &   & $\geq 180$ & $\geq 0.80$ & $73000$&$^{+1800}_{\minus 2400}$ & $70184$ \\[5pt]
VR-A2 & $5$ & $\geq 160$ & $\geq 0.85$ & $65000$&$^{+1800}_{\minus 2200}$ & $64985$ \\[5pt]
VR-A3 &   & $\geq 150$ & $\geq 0.90$ & $30000$&$^{+2100}_{\minus 1000}$ & $30360$ \\
\midrule[0.2pt]
VR-B1 &  & $\geq 120$ & $\geq 0.80$ & $80000$&$^{+2100}_{\minus 2800}$ & $80271$ \\[5pt]
VR-B2 & $6$ & $\geq 110$ & $\geq 0.85$ & $58000$&$^{+3900}_{\minus 1800}$ & $59997$ \\[5pt]
VR-B3 &   & $\geq 100$ & $\geq 0.90$ & $28000$&$^{+1000}_{\minus 2000}$ & $30212$ \\
\midrule[0.2pt]
VR-C1 &  & $\geq 180$ &    & $350$&$^{+37}_{\minus 72}$ & $372 $ \\[5pt]
VR-C2 & $\geq 7$ & $\geq 220$ & $\leq 0.60$ & $47$&$^{+6}_{\minus 10}$ & $35$ \\[5pt]
VR-C3 &  & $\geq 240$ &    & $18$&$^{+4}_{\minus 3}$ & $14$ \\
\midrule[0.2pt]
VR-D1 & $\geq 8$ & $\geq 180$ & $\leq 0.60$ & $23$&$^{+5}_{\minus 6}$ & $16$ \\
\bottomrule
\end{tabular}
}
\end{table}

\begin{table}[ht]
\caption{ \label{tab:cca-VR-bj-final}
VR definitions and yields, for the regions used to validate the background strategy with at least two $b$-tagged jets.
The common analysis selections on the $H_{\mathrm{T}}$, $p_{\mathrm{T}}(j_1)$
and trigger selection are also applied.
The uncertainties shown contain both the statistical and systematic uncertainties.
}
\centering{
\begin{tabular}{ c @{\hspace{3em}} C{3em} C{3em} C{4em}  @{\hspace{2.5em}} R{4.8em} @{\hspace{\arrayrulewidth}} L{3.2em}  r}
\toprule
& $n_{\mathrm{jets}}$ & $p_{\mathrm{T}}(j)$ [GeV] & $C$ & \multicolumn{2}{C{8em}}{Background Expectation} & \multicolumn{1}{C{3.5em}}{$\quad$Data} \\
\midrule
VR-A-bj & $5$      & $\geq 180$ & $\geq 0.85$ & $2100$ & $^{+600}_{-100}$ &  $1973$ \\[5pt]
VR-B-bj & $6$      & $\geq 120$ & $\geq 0.85$ & $3700$ & $^{+500}_{-300}$ &  $3425$ \\[5pt]
VR-C-bj & $\geq 7$ & $\geq 180$ & $\leq 0.60$ & $34$ & $^{+13}_{-6}$ &  $39$ \\[5pt]
VR-D-bj & $\geq 8$ & $\geq 160$ & $\leq 0.60$ & $8$ & $^{+6}_{-5}$ &  $6$ \\
\bottomrule
\end{tabular}
}
\end{table}
 
\clearpage

% End of text imported from the .//sections/Analysis_JC.tex input file

\subsection{Mass resonance method}

% The next lines are included from the .//sections/Analysis_ml.tex input file
The objective of the mass resonance method is to observe a resonance in the reconstructed candidate gluino mass spectrum.
In contrast to the jet counting method, which could have an excess from a variety of high energy contributions, this search would be an unambiguous sign of new physics from resonant production at the probed energy scale.
For the direct gluino decay model, the combinatorial issue of correctly identifying which jets should be matched to each gluino candidate is a significant problem and  the main objective of the design of the method.
A dedicated neural network (NN) is developed  to correctly group together the jets from each individual gluino candidate decay.
Machine-learning techniques were applied previously to the combinatorial assignment problem focusing on the reconstruction of standard model
processes~\cite{10.21468/SciPostPhys.12.5.178, PhysRevD.107.114029, PhysRevD.107.116019} and with less focus on beyond Standard Model scenarios~\cite{PhysRevD.106.016001} given the additional unknown of the exotic particle masses.
The invariant mass $m_{\tilde{g},i}$ of the two gluino candidates is built from the jets that are selected by the NN and the average of the two masses $m_{\mathrm{avg}} = \frac{1}{2}\left(m_{\tilde{g},1} + m_{\tilde{g},2}\right)$ is used as the key discriminating variable.
The method searches for a localised excess on the $m_{\mathrm{avg}}$ spectrum  where the background is estimated through a functional fit to a smooth decreasing spectrum.
 
Events used by the mass resonance method require  at least six jets with \pT above 100 \GeV, and $C \geq 0.9$. A second selection is defined requiring in addition at least one $b$-tagged jet, which is used to improve the sensitivity to the UDB model. Further to the previously introduced selections, which are used for the model-dependent interpretation, a set of model-independent SRs are defined using single bins in the invariant mass distribution with a width of 300 \GeV, and assume no signal contribution outside of the SR.

\subsubsection*{Jet assignment model}
 
A NN is built based on the attention mechanism~\cite{vaswani2017attention}, taking inspiration from the transformer model and implemented in
PyTorch~\cite{paszke2019pytorch}. The input to the network is the jet four-momentum of the leading eight signal jets, where jets are zero-padded if the event has less than eight jets. The first layer consists of an embedding block where each jet is mapped to a latent space. The embedded jets are passed to an encoder block consisting of a jet self-attention block, a gluino-candidate self-attention block, and a jet-candidate cross-attention block.
The outputs of the model are three scores per jet, representing the probability of the jet to originate from each of the two gluino candidates or a non-signal source such as initial-state radiation or pile-up.
The highest score per jet is used to assign jets in the event to each gluino or non-signal contribution.
 
The NN is trained using a categorical cross-entropy loss, where the jets are labelled based on $\Delta R < 0.4$ matching with MC generated partons from the gluino decay.
Only events with exactly three jets matched per gluino are used in the training, which represents approximately 50\% of the total available events.
A tighter preselection than introduced at the start of this section is applied to the training set to obtain a sample representative of the final kinematic selections while retaining large enough sample size, requiring at least six jets with \pT above 100~\GeV, and $C \geq 0.8$.
All the available signal masses are used in the training, combining both the UDS and UDB models.
The inclusion of all mass points was shown to mitigate the background sculpting by shifting the multijet average mass spectrum to roughly $650$~\GeV, below the start of the search window at $700$~\GeV.
The model is trained for $500,000$ steps with a warm-up phase which increases the learning rate linearly to $10^{-3}$ during the first 5\% of the training steps and is then decayed exponentially.
Hyper-parameters were tuned through a population scan and the model with lowest validation loss was retained~\cite{jaderberg2017population, liaw2018tune}.
 
\begin{figure}[t!]
\centering
\includegraphics[width=0.5\textwidth]{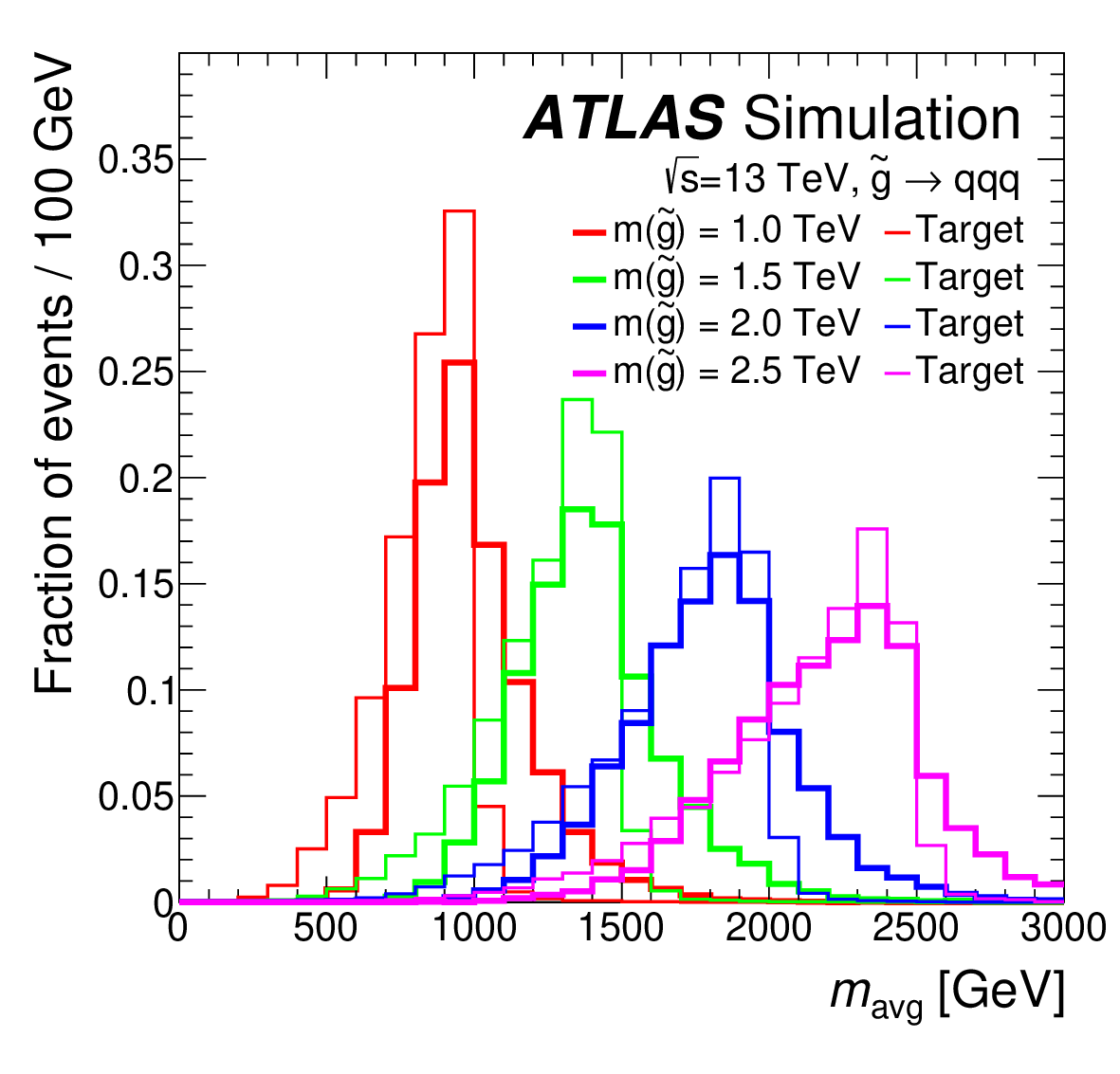}
\caption{Normalised average mass spectrum comparing the shapes of the reconstructed (solid) and target (light) distributions for different masses. The reconstructed distribution is produced using the NN assignments, whereas the target distribution is built assigning jets to gluinos based on their MC generated labels.}
\label{fig:predicted-mavg}
\end{figure}

The performance of the NN is illustrated in Figure~\ref{fig:predicted-mavg}. The reconstructed mass matches the target with a small loss in resolution as expected. Target signals show a low-mass tail, especially at higher masses, which originates from the restriction that exactly three jets are matched. This requirement misses additional final state radiation jets which are significant for the highest masses.
 
\subsubsection*{Background estimation}
 
The largest SM background for this search is non-resonant QCD processes, which result in multijet
systems with smoothly decreasing $m_{\mathrm{avg}}$ distributions.
To estimate this background, a parametric function is fit to the observed data distributions, which are binned in 100~\GeV-wide bins:
\begin{equation*}
f(x) = p_{1}\left(1-x\right)^{p_2}x^{p_3 + p_4 \ln{x}}  \,,
\end{equation*}
where $x=m_{\mathrm{avg}}/\sqrt{s}$ and $p_{1,2,3,4}$ are the fitted parameters.
This function is successfully used in a wide variety of resonance dijet and multijet searches by the CDF, CMS, and ATLAS experiments~\cite{CDF:2008ieg,EXOT-2010-01,CMS-EXO-11-015,EXOT-2013-11,EXOT-2015-02,CMS-EXO-16-032,EXOT-2020-15,EXOT-2019-03,CMS-EXO-21-010}.
For the background estimate, a three-parameter fit is used, where $p_{4}$ is set to zero, while the four-parameter fit is used to produce pseudodata to validate the fit strategy.
 
The fit to the background distribution uses a binned maximum-likelihood method that is implemented in the HistFitter framework~\cite{Baak:2014wma}. In background-only fits, the
signal strength is set to zero, while in the signal-plus-background fits, the signal strength is left as a free
parameter. The fit region is between $0.7$~\TeV and $3$~\TeV.
 
The data-driven background fitting procedure was validated with MC simulation
and a small 3.2 fb$^{-1}$ sample of data from 2015 with a loose selection of $C\geq0.7$ and at least six jets with $p_{\mathrm{T}}$ above $70$~\GeV
such that the number of events is similar to that of the full data sample with nominal selections.
The validity of the background model was tested by checking for a small $\chi^2/N_{\mathrm{D.O.F.}}$ and by performing `spurious signal tests' and `signal injection tests'.
 
The spurious signal test evaluates whether the fitting procedure is biased in a manner that will produce
a non-zero extracted signal when fitting a data sample with no true signal. This test is performed for the nominal
3-parameter fit function by performing a signal-plus-background fit to a pseudodata distribution that is generated from a background-only fit to the data
distribution with a 4-parameter function.
For each pseudodata distribution, the number of extracted signal events per signal model, $n_\text{spur}$, is determined.
To satisfy the spurious signal requirement, $n_\text{spur}$ is required to be less than 20\% of the nominal signal events
and the ratio of the number of spurious signal events to its statistical uncertainty, $n_\text{spur}/\sigma(n_\text{spur})$ is required to be less than 0.2.
The 3-parameter fit function passed the spurious signal test for all signal samples.
 
The signal injection test is performed to ensure that the background fit is able to extract a signal component
with the expected signal events. Simulated signal models are included together with the background template to form a pseudodata distribution.
The injected signals were extracted through the fit to pseudodata and confirmed that the extracted signal event yields were in agreement with the injected ones.
 
\begin{figure}[tb]
\centering
\subfigure[]{\includegraphics[width=0.49\textwidth]{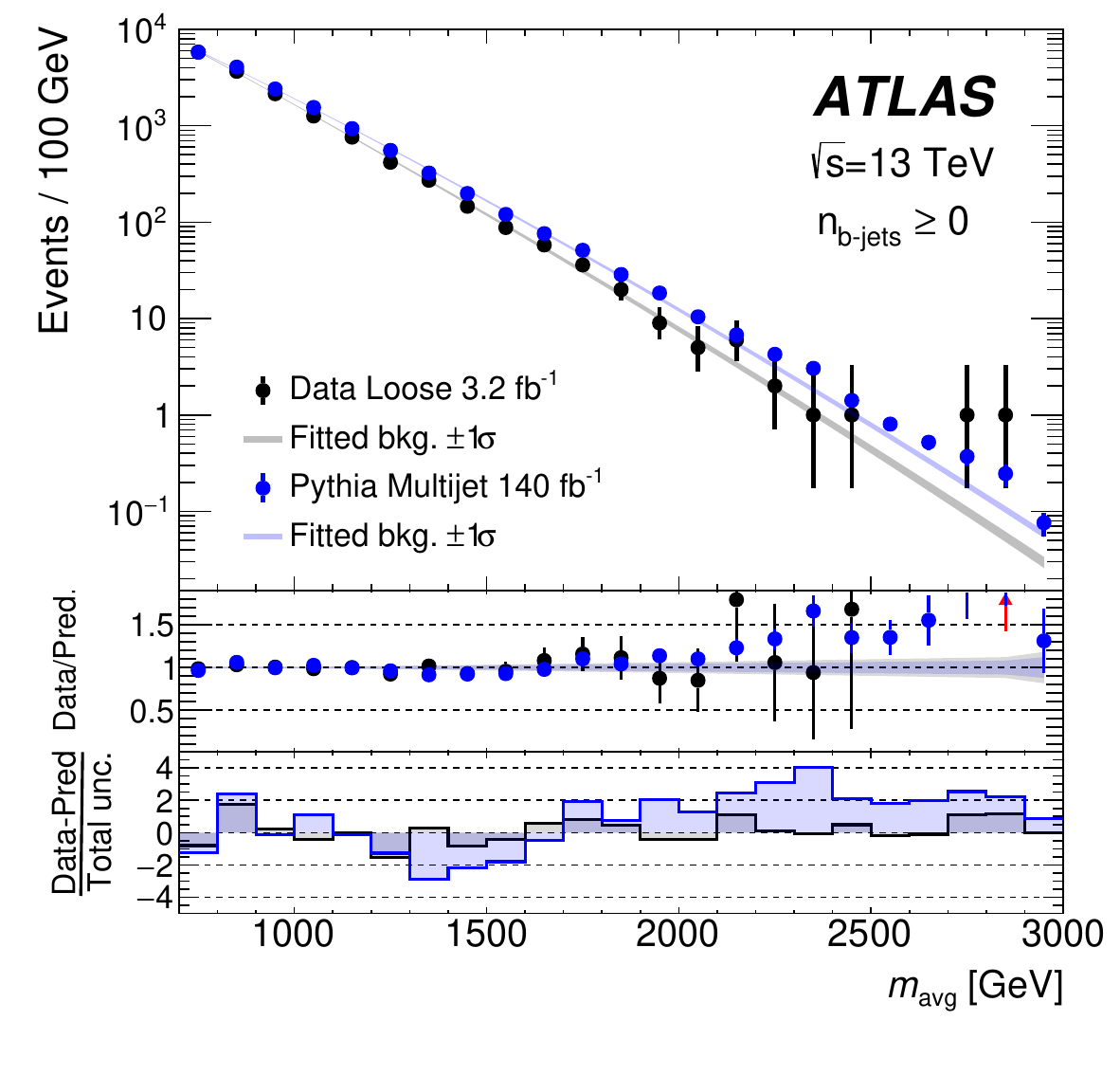}\label{fig:data15loose_uds}}
\subfigure[]{\includegraphics[width=0.49\textwidth]{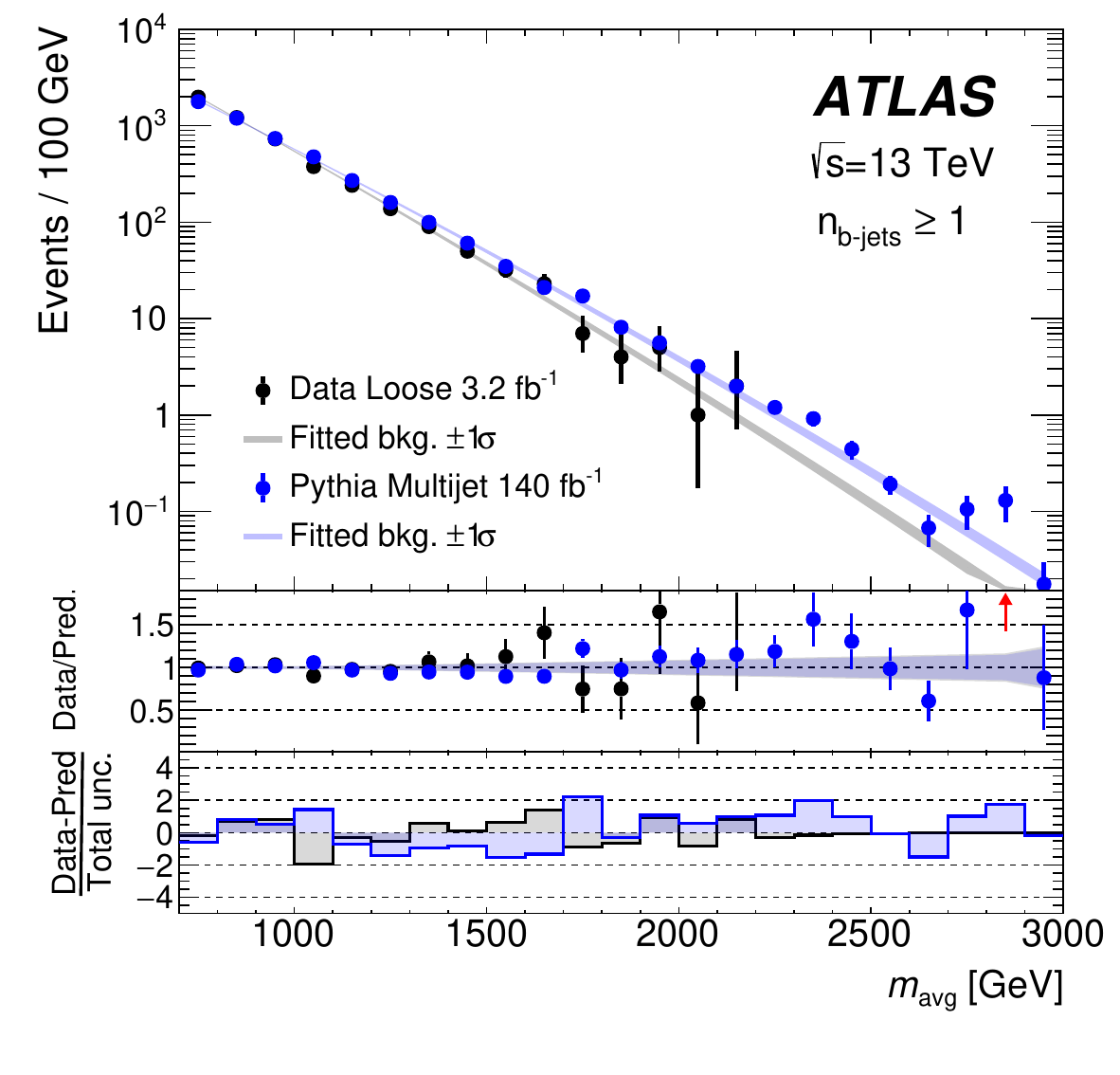}\label{fig:data15loose_udb}}
\caption{Observed data and the fit to the background model in the \subref{fig:data15loose_uds} nominal and \subref{fig:data15loose_udb} $b$-tagged regions using a loose selection and 3.2 fb$^{-1}$ of data, which matches roughly the number of events expected in the full selection and full data sample. The grey and blue bands present the combined statistical and systematic uncertainty in the background estimate for the data and MC fit functions respectively. The red arrow denotes points which lie above the range of the ratio plot.}
\label{fig:data15loose}
\end{figure}

Figure~\ref{fig:data15loose} shows the validation of the fit model in the 2015 data sample with loose selection and additionally with a multijet MC background sample scaled to 140\,fb$^{-1}$. The function with the best fit parameters has an acceptable agreement with both the MC and data.
 
When considering the model-independent SRs the background is estimated through a fit to the reconstructed average mass distribution excluding the 300 GeV wide signal region bin.
This is contrary to the model-dependent fits which are performed using the full average-mass distribution with 100~\GeV bins.

% End of text imported from the .//sections/Analysis_ml.tex input file

% End of text imported from the .//sections/Analysis.tex input file

\section{Systematic uncertainties}
% The next lines are included from the .//sections/Systematics.tex input file
Three categories of systematic uncertainties are considered in both the methods:
theoretical modelling uncertainties, experimental uncertainties, and uncertainties in the assumptions and methodologies used for the background estimate.
The statistical uncertainty due to the limited data sample is the dominant source of uncertainties for both the methods and for all of the mass range considered.
 
Modelling uncertainties related to the simulation of background events arise from missing higher orders in the simulation, as well as PDF and strong coupling constant $\alpha_{s}$ uncertainties.
They are included in the jet counting method as simulation is used to support the background prediction, but not in the mass resonance method.
The effect of these uncertainties in the multijet background yields, used to calculate the correction factors in the jet counting method, is evaluated through variations of the renormalisation and factorisation scale by factors of two, and variations of the shower tune, PDF and, $\alpha_{s}$ parameters within their uncertainties.
Additional uncertainties are included on the $t\bar{t}$ background comparing the nominal MC sample with other simulations:
an alternative matrix-element generator (\MGNLO) and an alternative parton shower (\HERWIG[7]).
The modelling of the QCD multijet background is the leading systematic uncertainty in the jet counting method.
In the SRs this uncertainty ranges from roughly 20\% up to almost 40\% in SR5.
 
Experimental uncertainties arise from the imperfect calibrations and the uncertainties in the reconstructed objects used in the search. The leading experimental uncertainties arise from the jet energy scale and jet energy resolution. Uncertainties in the pile-up modelling, suppression of pile-up jets, $b$-tagging efficiencies and mis-tagging rates are included but have a negligible impact on the sensitivity. The impact of experimental uncertainties is also considered for signal samples and correlated with the background variation.
The uncertainty in the combined 2015--2018 integrated luminosity is 0.83\%~\cite{DAPR-2021-01}, obtained using the LUCID-2 detector~\cite{LUCID2} for the primary luminosity measurements, complemented by measurements using the inner detector and calorimeters.
 
Dedicated additional uncertainties due the background estimate methodology are included.
In the jet counting method the observed level of agreement in the VRs is used to derive an uncertainty due to possible imperfections in the method.
The level of disagreement is below one standard deviation in all VRs except for VR-B3.
As previously mentioned, an additional non-closure systematic of 5\% is added to all SRs with $C>0.9$, which is required to cover the maximum non-closure in the VRs.
 
In the mass resonance method a spurious signal uncertainty is derived by fitting the distribution obtained from a 4-parameter background fit to a signal plus background hypothesis using the nominal 3-parameter model.
The size of the fitted signal is included as a systematic uncertainty due to possible limitations of the 3-parameter model to capture the correct background distribution.
The uncertainty varies from roughly 300 events for a reconstructed average mass of 900~\GeV to five events at 2500~\GeV.
The spurious signal uncertainty is the largest systematic in the mass resonance method.
However, the overall largest uncertainty is the statistical uncertainty from the background prediction.

% End of text imported from the .//sections/Systematics.tex input file

\section{Results and interpretation}
\begin{figure}[htbp]
\centering
\subfigure[]{\includegraphics[width=0.49\textwidth]{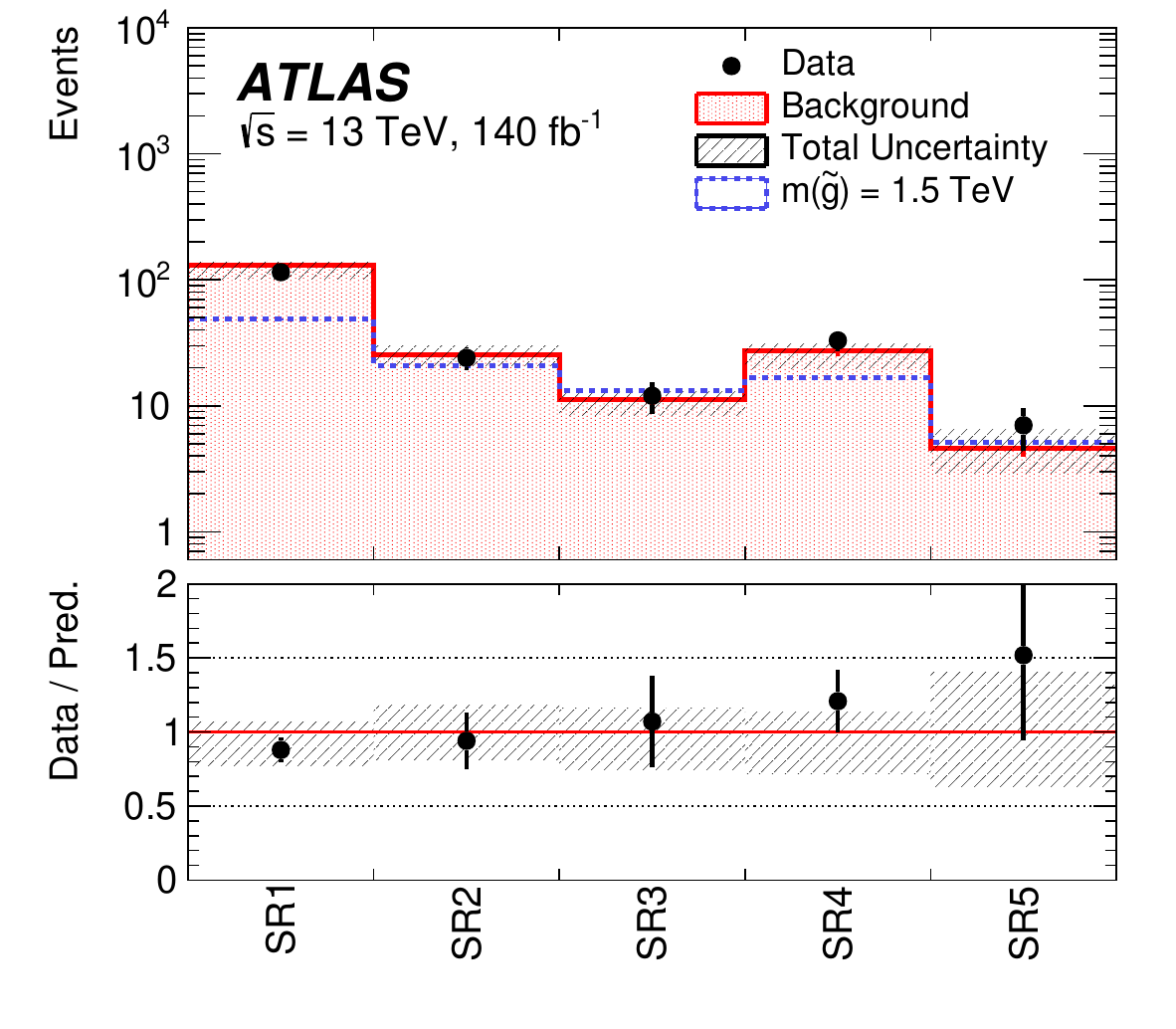}\label{fig:SRs-cca_a}}
\subfigure[]{\includegraphics[width=0.49\textwidth]{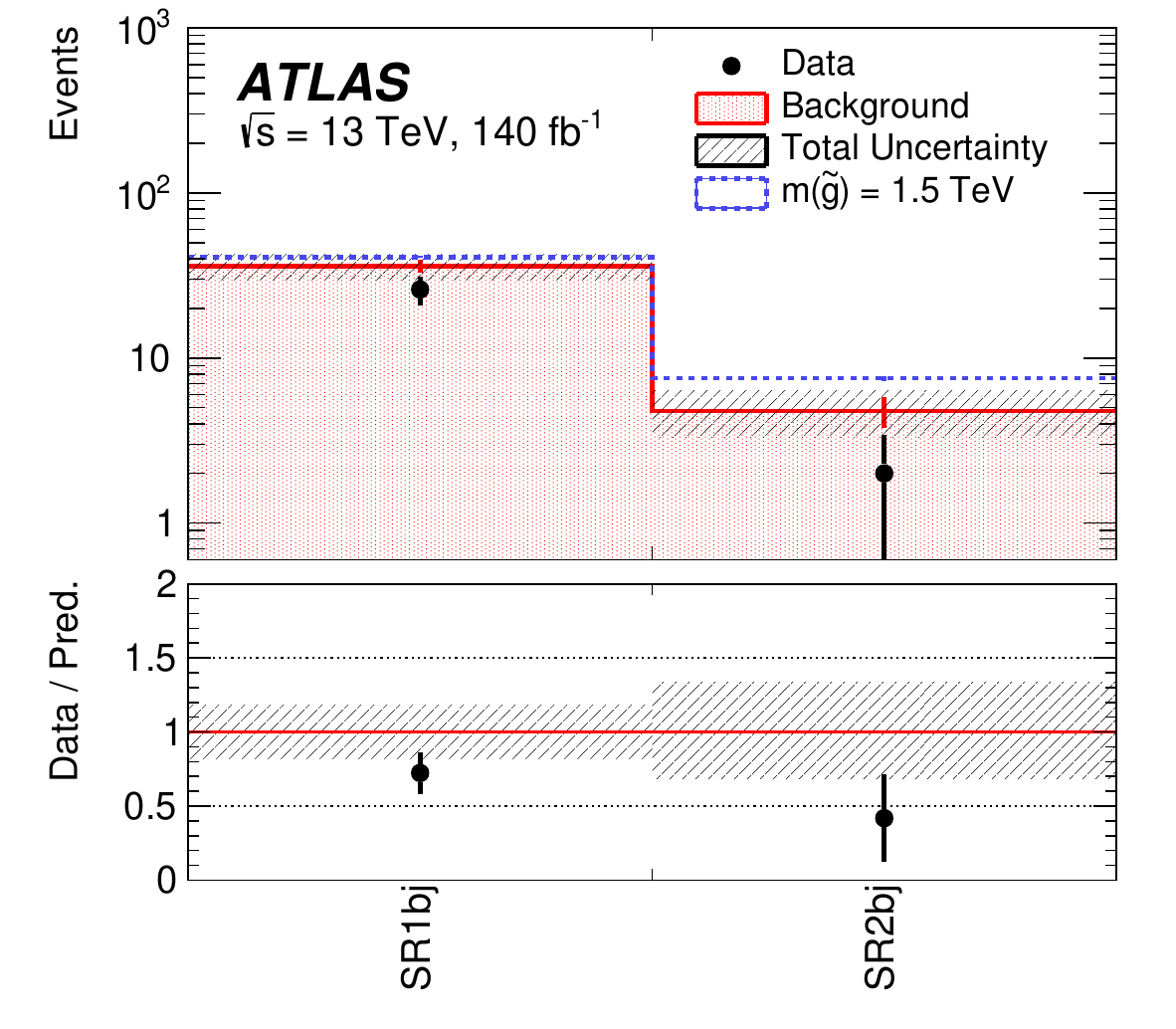}\label{fig:SRs-cca_b}}
\caption{
Observed and predicted event yields in the signal regions of the jet counting method for the \subref{fig:SRs-cca_a} $b$-tagging inclusive,
and \subref{fig:SRs-cca_b} $\geq 2$ $b$-tags regions.
The bottom panel presents the ratio of data to the background prediction.
The hatched pattern represents the combined statistical and systematic uncertainty in the background estimate.
}
\label{fig:SRs-cca}
\end{figure}

\begin{figure}[htbp]
\centering
\subfigure[]{\includegraphics[width=0.49\textwidth]{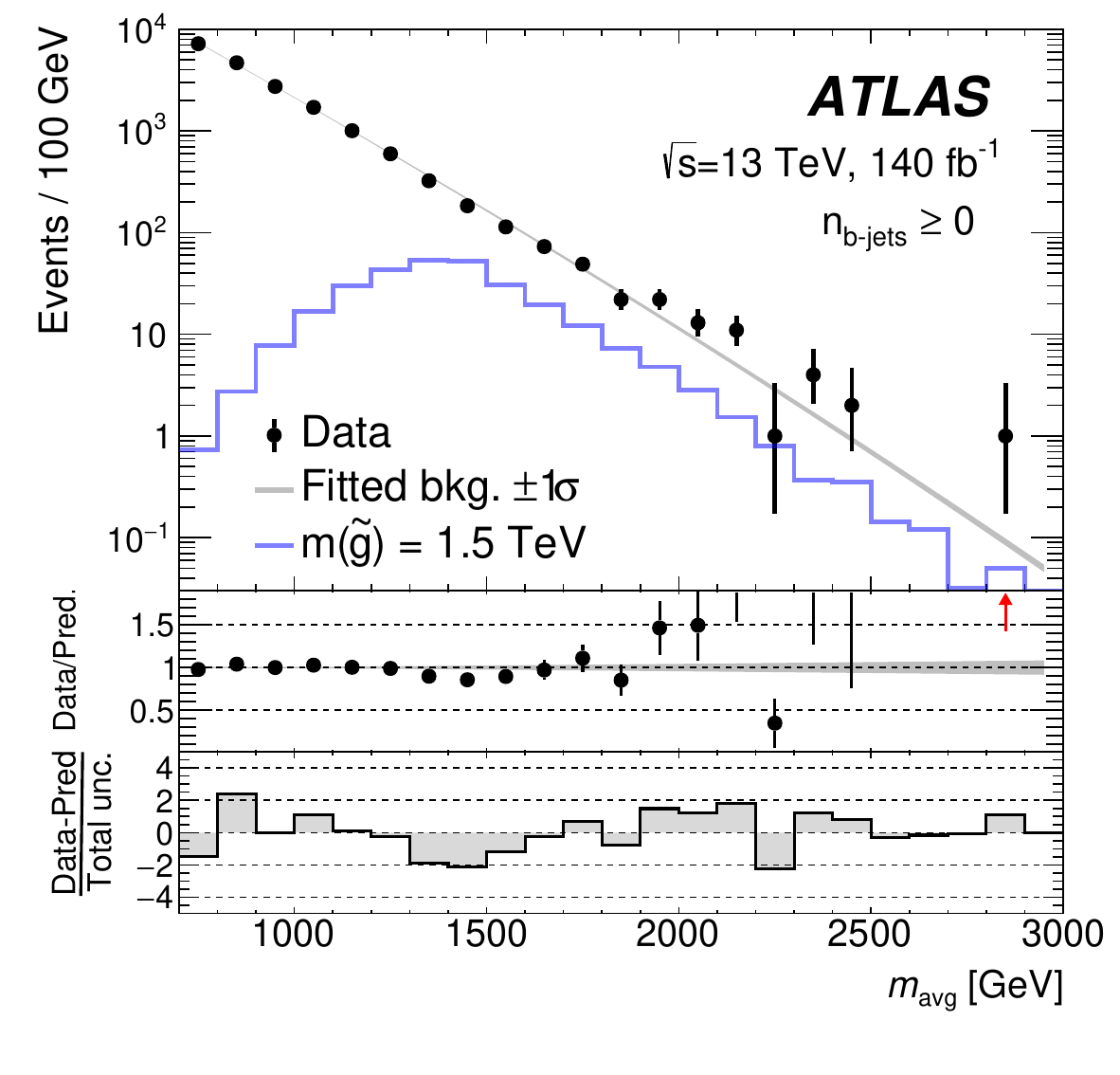}\label{fig:SRs-ml_a}}
\subfigure[]{\includegraphics[width=0.49\textwidth]{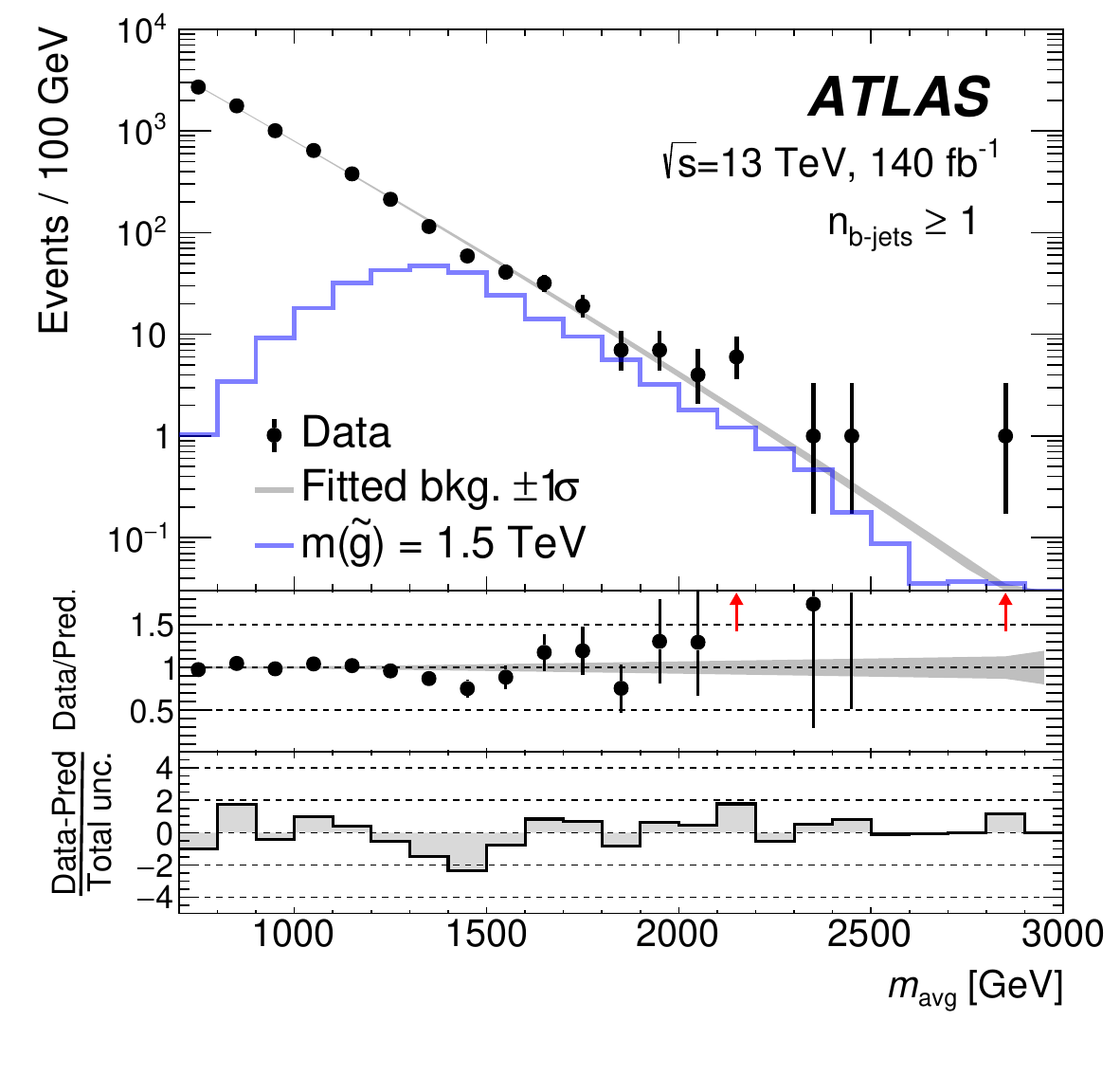}\label{fig:SRs-ml_b}}
\caption{
Background-only fits to the reconstructed average mass spectrum of the candidate gluinos, in \subref{fig:SRs-ml_a} the nominal and \subref{fig:SRs-ml_b} $b$-tagged regions of the mass resonance method.
The grey bands include both the statistical and systematic uncertainties.
The red arrow denotes points that lie above the range of the ratio plot.
}
\label{fig:SRs-ml}
\end{figure}

% The next lines are included from the .//sections/Results.tex input file
The observed data event yields and the corresponding estimates for the backgrounds in the SRs are shown in Figures~\ref{fig:SRs-cca} and~\ref{fig:SRs-ml} for the jet counting and mass resonance methods.
No significant excess of data over the expected event yields is observed in any of the SRs.
 
The profile likelihood-ratio test statistic~\cite{Cowan:2010js} is used to establish 95\% confidence intervals using the CL$_\text{s}$ prescription~\cite{Read:2002hq}. The uncertainties introduced in the previous section are included as nuisance parameters described by a Gaussian distribution. The asymptotic approximation~\cite{Cowan:2010js} of the CL$_\text{s}$ is used for all statistical tests except for the high-mass model-independent bins of the mass resonance method, where the expected number of events is small. In this case the CL$_\text{s}$ is computed using pseudo-experiments generated from simulated events.
The asymptotic approximation is validated for other regions with moderately small yields for both the methods using toys. Upper limits on the product of cross-section, acceptance, and efficiency are shown in Table~\ref{fig:cca-unblinded-discoveryModeUpperLimitTable} for the jet counting method, and in Tables~\ref{fig:ml-unblinded-discoveryModeUpperLimitTable-UDS} and~\ref{fig:ml-unblinded-discoveryModeUpperLimitTable-UDB} for the mass resonance method. The upper limits range from 7.9 fb to 0.03 fb, depending on the signal region considered.

% The next lines are included from the .//tables/cca/UpperLimitTableCombined_CCA_Rounded.tex input file
 
\begin{table}
\centering
\caption[Model independent upper limits CCA]{
The upper limit table for the signal regions for the jet counting method.
Left to right: 95\% CL upper limits on the visible cross-section
($\langle\epsilon\sigma\rangle_{\mathrm{obs}}^{95}$) and on the number of
signal events ($S_{\mathrm{obs}}^{95}$ ).
The third column ($S_{\mathrm{exp}}^{95}$) shows the 95\% CL upper limit
on the number of signal events, given the expected number
(and $\pm 1\sigma$ excursions on the expectation) of background events.
The last two columns
indicate the $CL_\mathrm{B}$ value, i.e., the confidence level observed for
the background-only hypothesis,
and the discovery $p$-value ($p_0$), with its corresponding Gaussian significance (Z).
The $p_0$ measures the compatibility of the observed data with the background-only (zero signal strength) hypothesis, relative to fluctuations of the background.
Larger values indicate greater relative compatibility.
In signal regions with a deficit relative to the nominal background prediction, the $p_0$ value is capped at 0.50.
}
\setlength{\tabcolsep}{0.0pc}
\begin{tabular*}{\textwidth}{@{\extracolsep{\fill}}lcr r@{\hspace{-36pt}}l cc}
\noalign{\smallskip}\hline\noalign{\smallskip}
Signal region & $\langle\epsilon{\mathrm \sigma}\rangle_{\mathrm{obs}}^{95}$[fb]  &  $S_{\mathrm{obs}}^{95}$  &  \multicolumn{2}{l}{ $\quad S_{\mathrm{exp}}^{95}$ } & $CL_{\mathrm{B}}$ & $p_{0}$ ($Z$)  \\
\noalign{\smallskip}\hline\noalign{\smallskip}
SR1    & $0.32$ &  $45$ & $ { 57 }$ & $^{ +18 }_{ \minus 14 }$ & $0.51$ & $ 0.50$~$(0.00)$ \\[3pt]
SR2    & $0.09$ &  $13$ & $ { 14.1 }$ & $^{ +5.7 }_{ \minus 4.1 }$ & $0.56$ & $ 0.50$~$(0.00)$ \\[3pt]
SR3    & $0.07$ &  $10$ & $ { 9.5 }$ & $^{ +4.1 }_{ \minus 2.7 }$ & $0.52$ & $ 0.42$~$(0.20)$ \\[3pt]
SR4    & $0.16$ &  $22$ & $ { 17.4 }$ & $^{ +6.5 }_{ \minus 4.7 }$ & $0.26$ & $ 0.21$~$(0.79)$ \\[3pt]
SR5    & $0.07$ &  $9.4$ & $ { 7.4 }$ & $^{ +3.6 }_{ \minus 2.4 }$ & $0.42$ & $ 0.32$~$(0.46)$ \\[3pt]
SR1bj  & $0.08$ &  $11$ & $ { 17.0 }$ & $^{ +6.9 }_{ \minus 4.8 }$ & $0.55$ & $ 0.50$~$(0.00)$ \\[3pt]
SR2bj  & $0.03$ &  $4.4$ & $ { 6.6 }$ & $^{ +2.9 }_{ \minus 1.9 }$ & $0.66$ & $ 0.50$~$(0.00)$ \\%
 
\noalign{\smallskip}\hline\noalign{\smallskip}
\end{tabular*}
\label{fig:cca-unblinded-discoveryModeUpperLimitTable}
\end{table}

% End of text imported from the .//tables/cca/UpperLimitTableCombined_CCA_Rounded.tex input file

% The next lines are included from the .//tables/ml/unblinding/UpperLimitTableCombined_UDS_Rounded.tex input file
\begin{table}
\centering
\caption[Breakdown of upper limits.]{
The upper limit table for the $\geq 0$ $b$-tagged jets region of the mass resonance method.
Left to right: 95\% CL upper limits on the visible cross-section
($\langle\epsilon\sigma\rangle_{\mathrm{obs}}^{95}$) and on the number of
signal events ($S_{\mathrm{obs}}^{95}$ ).
The third column ($S_{\mathrm{exp}}^{95}$) shows the 95\% CL upper limit
on the number of signal events, given the expected number
(and $\pm 1\sigma$ excursions on the expectation) of background events.
The last two columns
indicate the $CL_\mathrm{B}$ value, i.e., the confidence level observed for
the background-only hypothesis,
and the discovery $p$-value ($p_0$), with its corresponding Gaussian significance (Z).
The $p_0$ measures the compatibility of the observed data with the background-only (zero signal strength) hypothesis, relative to fluctuations of the background.
Larger values indicate greater relative compatibility.
In signal regions with a deficit relative to the nominal background prediction, the $p_0$ value is capped at 0.50.
}
\setlength{\tabcolsep}{0.0pc}
\begin{tabular*}{\textwidth}{@{\extracolsep{\fill}}crr r @{\hspace{-33pt}} l cr}
\noalign{\smallskip}\hline\noalign{\smallskip}
m$_\mathrm{avg}$ range [\GeV] & $\langle\epsilon{\mathrm \sigma}\rangle_{\mathrm{obs}}^{95}$[fb]  &  $S_{\mathrm{obs}}^{95}$  &  \multicolumn{2}{l}{ $\quad S_{\mathrm{exp}}^{95}$ } & $CL_{\mathrm{B}}$ & $p_{0}$ ($Z$)  \\
\noalign{\smallskip}\hline\noalign{\smallskip}
700 -- 1000  & $7.3$ &  $1000$ & $ { 1300 }$ & $^{ +460 }_{ \minus 300 }$ & $0.22$ & $ 0.50$~$(0.00)$ \\[3pt]
800 -- 1100  & $5.7$ &  $800$ & $ { 360 }$ & $^{ +150 }_{ \minus 49 }$ & $0.99$ & $ 0.01$~$(2.5)$ \\[3pt]
900 -- 1200  & $2.1$ &  $290$ & $ { 210 }$ & $^{ +88 }_{ \minus 25 }$ & $0.81$ & $ 0.18$~$(0.91)$ \\[3pt]
1000 -- 1300  & $1.5$ &  $210$ & $ { 160 }$ & $^{ +50 }_{ \minus 34 }$ & $0.80$ & $ 0.18$~$(0.90)$ \\[3pt]
1100 -- 1400  & $0.54$ &  $76$ & $ { 120 }$ & $^{ +45 }_{ \minus 30 }$ & $0.09$ & $ 0.50$~$(0.00)$ \\[3pt]
1200 -- 1500  & $0.27$ &  $37$ & $ { 85 }$ & $^{ +33 }_{ \minus 24 }$ & $0.00$ & $ 0.50$~$(0.00)$ \\[3pt]
1300 -- 1600  & $0.16$ &  $23$ & $ { 63 }$ & $^{ +37 }_{ \minus 18 }$ & $0.00$ & $ 0.50$~$(0.00)$ \\[3pt]
1400 -- 1700  & $0.16$ &  $22$ & $ { 47 }$ & $^{ +19 }_{ \minus 13 }$ & $0.00$ & $ 0.50$~$(0.00)$ \\[3pt]
1500 -- 1800  & $0.24$ &  $33$ & $ { 39 }$ & $^{ +16 }_{ \minus 10 }$ & $0.25$ & $ 0.50$~$(0.00)$ \\[3pt]
1600 -- 1900  & $0.26$ &  $37$ & $ { 38 }$ & $^{ +15 }_{ \minus 10 }$ & $0.47$ & $ 0.50$~$(0.00)$ \\[3pt]
1700 -- 2000  & $0.30$ &  $42$ & $ { 34 }$ & $^{ +12 }_{ \minus 7 }$ & $0.71$ & $ 0.29$~$(0.55)$ \\[3pt]
1800 -- 2100  & $0.25$ &  $35$ & $ { 28 }$ & $^{ +12 }_{ \minus 8 }$ & $0.72$ & $ 0.28$~$(0.57)$ \\[3pt]
1900 -- 2200  & $0.29$ &  $41$ & $ { 25 }$ & $^{ +11 }_{ \minus 4 }$ & $0.93$ & $ 0.06$~$(1.5)$ \\[3pt]
2000 -- 2300  & $0.19$ &  $27$ & $ { 21.5 }$ & $^{ +7.6 }_{ \minus 4.4 }$ & $0.78$ & $ 0.19$~$(0.89)$ \\[3pt]
2100 -- 2400  & $0.15$ &  $21$ & $ { 15.5 }$ & $^{ +6.2 }_{ \minus 2.3 }$ & $0.74$ & $ 0.20$~$(0.84)$ \\[3pt]
2200 -- 2500  & $0.08$ &  $11$ & $ { 10.5 }$ & $^{ +3.2 }_{ \minus 1.9 }$ & $0.57$ & $ 0.40$~$(0.26)$ \\[3pt]
2300 -- 2600  & $0.08$ &  $11$ & $ { 9.2 }$ & $^{ +3.9 }_{ \minus 1.2 }$ & $0.66$ & $ 0.27$~$(0.61)$ \\[3pt]
2400 -- 2700  & $0.05$ &  $6.9$ & $ { 6.8 }$ & $^{ +2.1 }_{ \minus 1.4 }$ & $0.51$ & $ 0.48$~$(0.05)$ \\[3pt]
2500 -- 2800  & $0.02$ &  $2.3$ & $ { 3.1 }$ & $^{ +2.1 }_{ \minus 1.2 }$ & $0.26$ & $ 0.50$~$(0.01)$ \\[3pt]
2600 -- 2900  & $0.04$ &  $5.3$ & $ { 5.2 }$ & $^{ +2.2 }_{ \minus 1.3 }$ & $0.52$ & $ 0.46$~$(0.10)$ \\[3pt]
2700 -- 3000  & $0.06$ &  $8.3$ & $ { 8.2 }$ & $^{ +0.4 }_{ \minus 0.7 }$ & $0.53$ & $ 0.44$~$(0.16)$ \\%
\noalign{\smallskip}\hline\noalign{\smallskip}
\end{tabular*}
\label{fig:ml-unblinded-discoveryModeUpperLimitTable-UDS}
\end{table}

% End of text imported from the .//tables/ml/unblinding/UpperLimitTableCombined_UDS_Rounded.tex input file

% The next lines are included from the .//tables/ml/unblinding/UpperLimitTableCombined_UDB_Rounded.tex input file
\begin{table}
\centering
\caption[Breakdown of upper limits.]{
The upper limit table for the $\geq 1$ $b$-tagged jets region of the mass resonance method. Left to right: 95\% CL upper limits on the visible cross-section
($\langle\epsilon\sigma\rangle_{\mathrm{obs}}^{95}$) and on the number of
signal events ($S_{\mathrm{obs}}^{95}$ ).
The third column ($S_{\mathrm{exp}}^{95}$) shows the 95\% CL upper limit
on the number of signal events, given the expected number
(and $\pm 1\sigma$ excursions on the expectation) of background events.
The last two columns
indicate the $CL_\mathrm{B}$ value, i.e., the confidence level observed for
the background-only hypothesis,
and the discovery $p$-value ($p_0$), with its corresponding Gaussian significance (Z).
The $p_0$ measures the compatibility of the observed data with the background-only (zero signal strength) hypothesis, relative to fluctuations of the background.
Larger values indicate greater relative compatibility.
In signal regions with a deficit relative to the nominal background prediction, the $p_0$ value is capped at 0.50.
}
\setlength{\tabcolsep}{0.0pc}
\begin{tabular*}{\textwidth}{@{\extracolsep{\fill}}crr r @{\hspace{-33pt}} l cr}
\noalign{\smallskip}\hline\noalign{\smallskip}
m$_\mathrm{avg}$ range [\GeV] & $\langle\epsilon{\mathrm \sigma}\rangle_{\mathrm{obs}}^{95}$[fb]  &  $S_{\mathrm{obs}}^{95}$  &  \multicolumn{2}{l}{ $\quad S_{\mathrm{exp}}^{95}$ } & $CL_{\mathrm{B}}$ & $p_{0}$ ($Z$)  \\
\noalign{\smallskip}\hline\noalign{\smallskip}
700 -- 1000  & $5.7$ &  $800$ & $ { 960 }$ & $^{ +330 }_{ -240 }$ & $0.31$ & $ 0.50$~$(0.00)$ \\[3pt]
800 --1100  & $3.3$ &  $460$ & $ { 320 }$ & $^{ +100 }_{ -65 }$ & $0.89$ & $ 0.11$~$(1.20)$ \\[3pt]
900 -- 1200  & $1.1$ &  $150$ & $ { 130 }$ & $^{ +38 }_{ -31 }$ & $0.74$ & $ 0.24$~$(0.71)$ \\[3pt]
1000 -- 1300  & $0.92$ &  $130$ & $ { 92 }$ & $^{ +40 }_{ -13 }$ & $0.81$ & $ 0.18$~$(0.91)$ \\[3pt]
1100 -- 1400  & $0.36$ &  $51$ & $ { 70 }$ & $^{ +27 }_{ -20 }$ & $0.17$ & $ 0.50$~$(0.00)$ \\[3pt]
1200 -- 1500  & $0.16$ &  $23$ & $ { 52 }$ & $^{ +21 }_{ -15 }$ & $0.00$ & $ 0.50$~$(0.00)$ \\[3pt]
1300 -- 1600  & $0.11$ &  $16$ & $ { 39 }$ & $^{ +15 }_{ -11 }$ & $0.00$ & $ 0.50$~$(0.00)$ \\[3pt]
1400 -- 1700  & $0.12$ &  $17$ & $ { 29 }$ & $^{ +12 }_{ -8 }$ & $0.04$ & $ 0.50$~$(0.00)$ \\[3pt]
1500 -- 1800  & $0.20$ &  $27$ & $ { 25 }$ & $^{ +25 }_{ -7 }$ & $0.61$ & $ 0.38$~$(0.29)$ \\[3pt]
1600 -- 1900  & $0.25$ &  $35$ & $ { 30 }$ & $^{ +10 }_{ -7 }$ & $0.68$ & $ 0.45$~$(0.13)$ \\[3pt]
1700 -- 2000  & $0.21$ &  $30$ & $ { 28 }$ & $^{ +10 }_{ -8 }$ & $0.58$ & $ 0.42$~$(0.20)$ \\[3pt]
1800 -- 2100  & $0.17$ &  $24$ & $ { 24.0 }$ & $^{ +5.9 }_{ -6.2 }$ & $0.51$ & $ 0.49$~$(0.03)$ \\[3pt]
1900 -- 2200  & $0.18$ &  $25$ & $ { 21.6 }$ & $^{ +5.9 }_{ -5.8 }$ & $0.71$ & $ 0.26$~$(0.65)$ \\[3pt]
2000 -- 2300  & $0.13$ &  $18$ & $ { 17.1 }$ & $^{ +5.3 }_{ -2.1 }$ & $0.63$ & $ 0.32$~$(0.47)$ \\[3pt]
2100 -- 2400  & $0.10$ &  $13$ & $ { 12.4 }$ & $^{ +3.3 }_{ -2.6 }$ & $0.63$ & $ 0.30$~$(0.51)$ \\[3pt]
2200 -- 2500  & $0.05$ &  $6.4$ & $ { 6.4 }$ & $^{ +2.5 }_{ -1.5 }$ & $0.50$ & $ 0.50$~$(0.00)$ \\[3pt]
2300 -- 2600  & $0.05$ &  $6.8$ & $ { 6.7 }$ & $^{ +2.6 }_{ -0.8 }$ & $0.54$ & $ 0.42$~$(0.20)$ \\[3pt]
2400 -- 2700  & $0.03$ &  $4.0$ & $ { 3.9 }$ & $^{ +2.2 }_{ -1.2 }$ & $0.52$ & $ 0.45$~$(0.14)$ \\[3pt]
2500 -- 2800  & $0.01$ &  $2.0$ & $ { 2.1 }$ & $^{ +1.8 }_{ -0.9 }$ & $0.47$ & $ 0.49$~$(0.02)$ \\[3pt]
2600 -- 2900  & $0.04$ &  $5.4$ & $ { 5.3 }$ & $^{ +2.2 }_{ -1.3 }$ & $0.53$ & $ 0.43$~$(0.19)$ \\[3pt]
2700 -- 3000  & $0.04$ &  $6.1$ & $ { 6.0 }$ & $^{ +2.3 }_{ -0.6 }$ & $0.53$ & $ 0.42$~$(0.20)$ \\%
\noalign{\smallskip}\hline\noalign{\smallskip}
\end{tabular*}
\label{fig:ml-unblinded-discoveryModeUpperLimitTable-UDB}
\end{table}

% End of text imported from the .//tables/ml/unblinding/UpperLimitTableCombined_UDB_Rounded.tex input file

\begin{figure}[htbp!]
\centering
\subfigure[]{\includegraphics[width=0.49\textwidth]{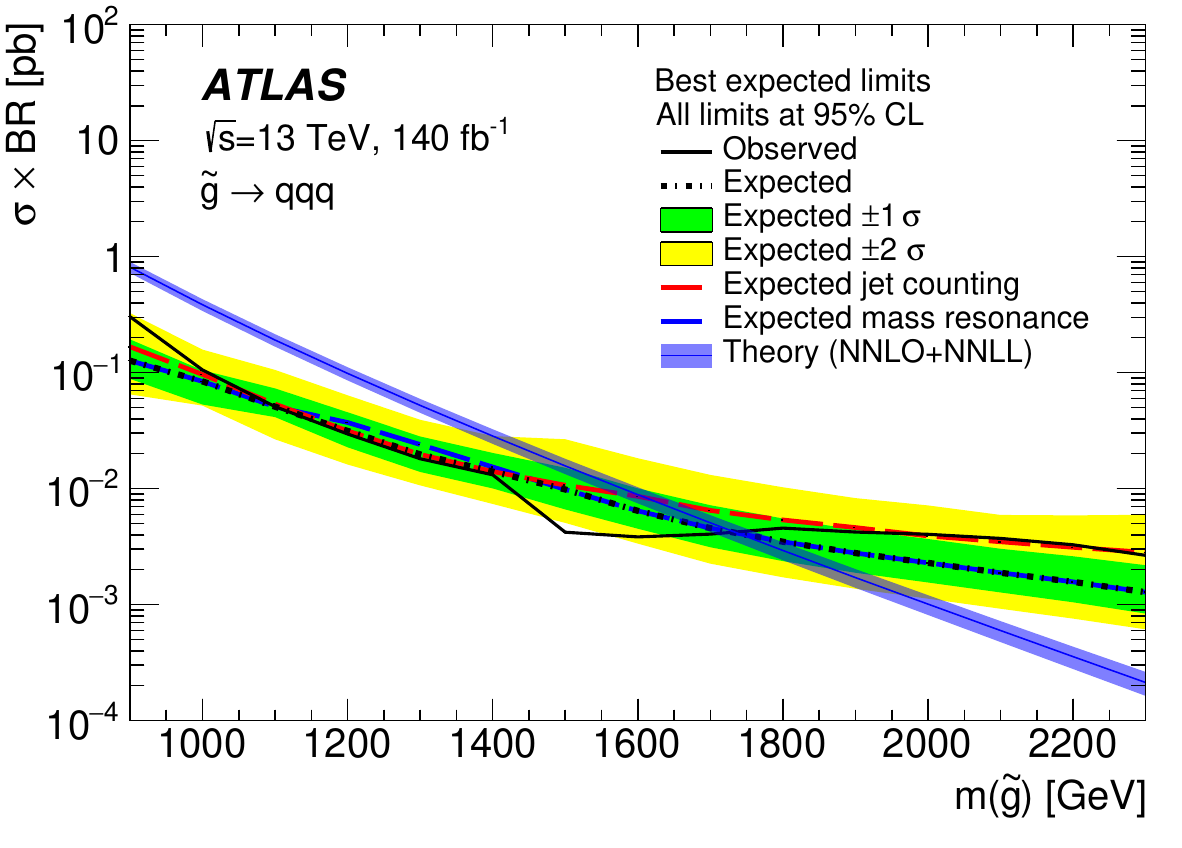}\label{fig:cca-obs-lim-2t3_BestExpected_a}} 
\subfigure[]{\includegraphics[width=0.49\textwidth]{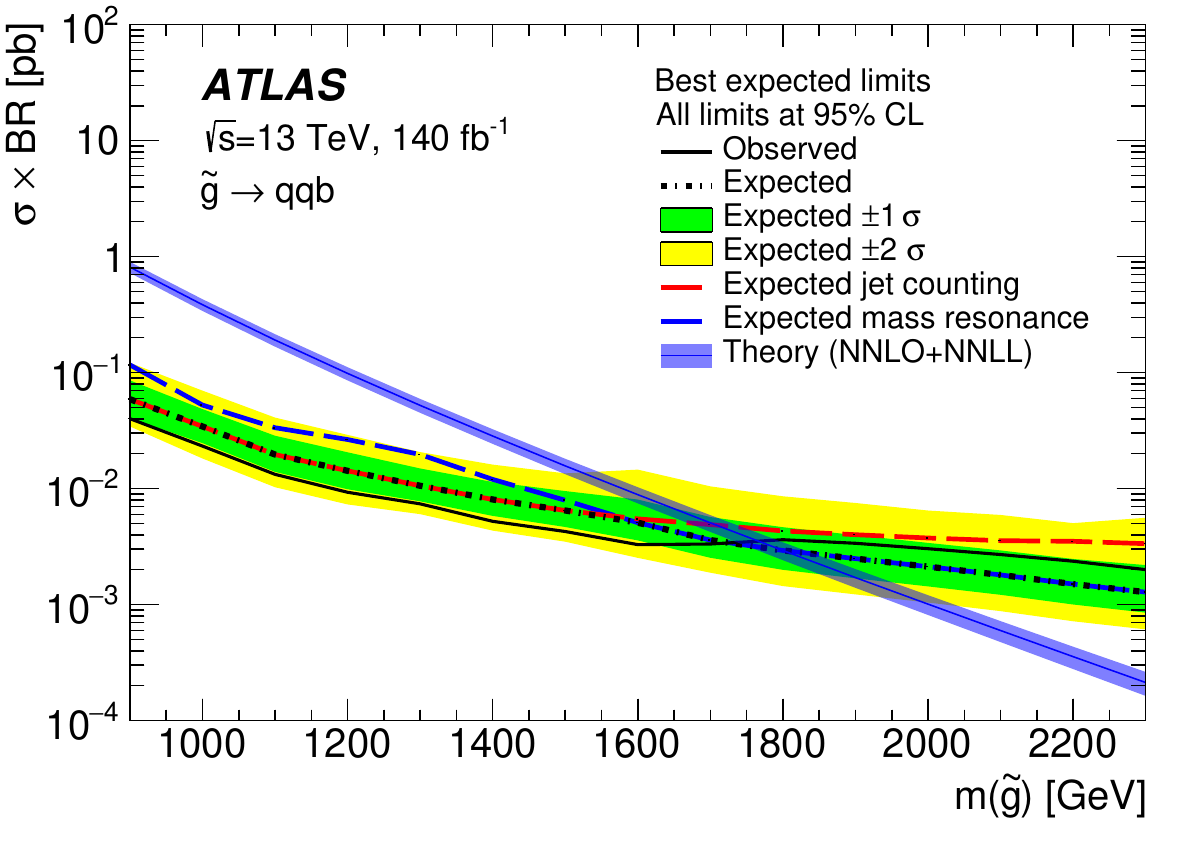}\label{fig:cca-obs-lim-2t3_BestExpected_b}} 
 
\caption{
Observed and expected 95\% CL upper limits on the signal cross-section times branching ratio ($\sigma \;\times$ BR) as a function of the gluino mass for the gluino direct decay model with \subref{fig:cca-obs-lim-2t3_BestExpected_a} \UDS and \subref{fig:cca-obs-lim-2t3_BestExpected_b} \UDB decays.
The expected limits for the jet counting and mass resonance methods are shown in red and blue, respectively. The best expected limit per mass point between the methods is chosen (dashed black) and corresponding observed limit reported (solid black).
The green and yellow bands around the expected limit correspond to the $\pm 1 \sigma$ and $\pm 2 \sigma$ variations including both the systematic and statistical uncertainties, respectively.
The theoretical prediction is also shown, with the uncertainties in the prediction shown as a coloured band.
}
\label{fig:cca-obs-lim-2t3_BestExpected}
\end{figure}

\begin{figure}[htpb!]
\centering
\subfigure[]{\includegraphics[width=0.49\textwidth]{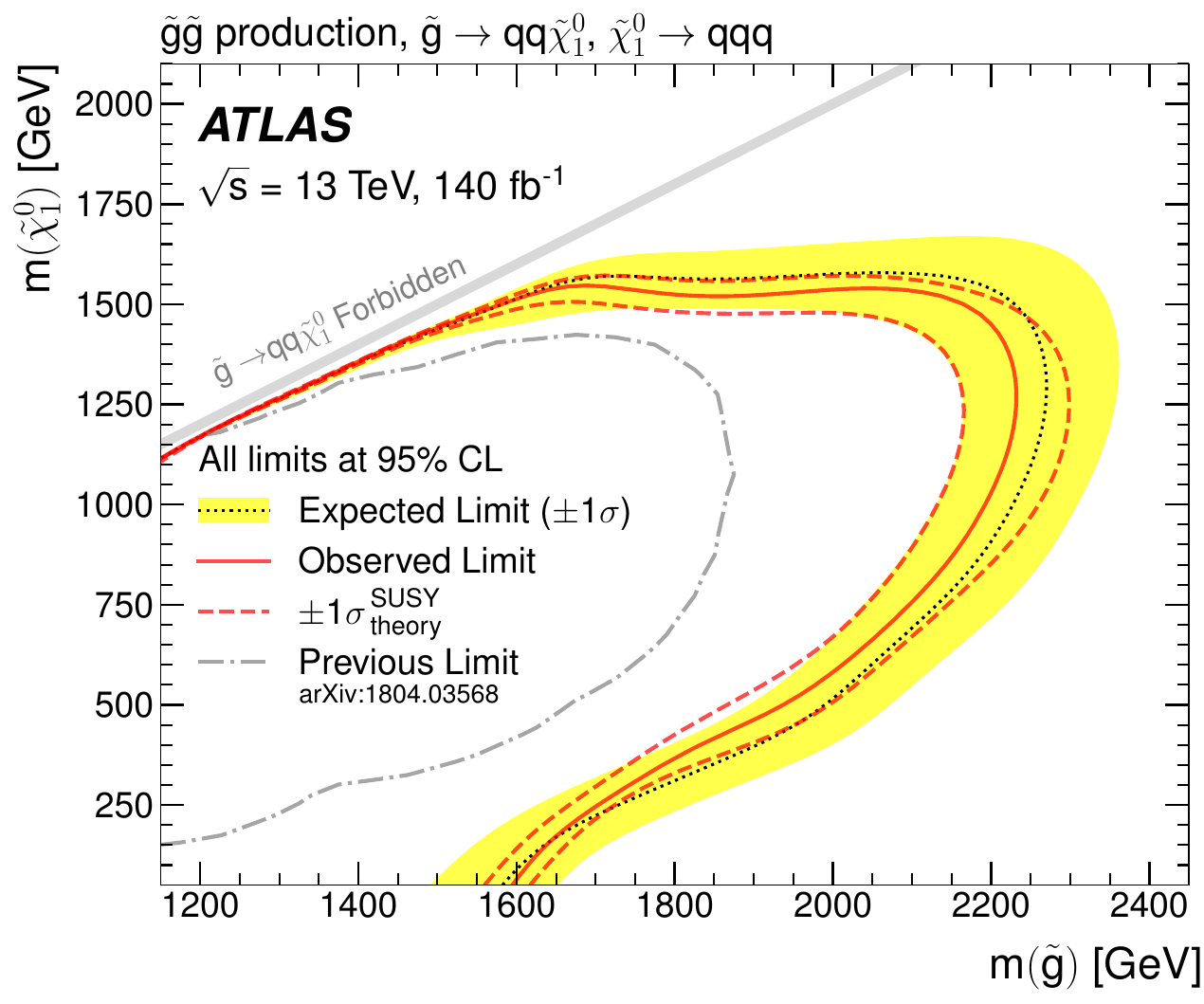} \label{fig:cca-obs-lim-2t5_a}}
\subfigure[]{\includegraphics[width=0.49\textwidth]{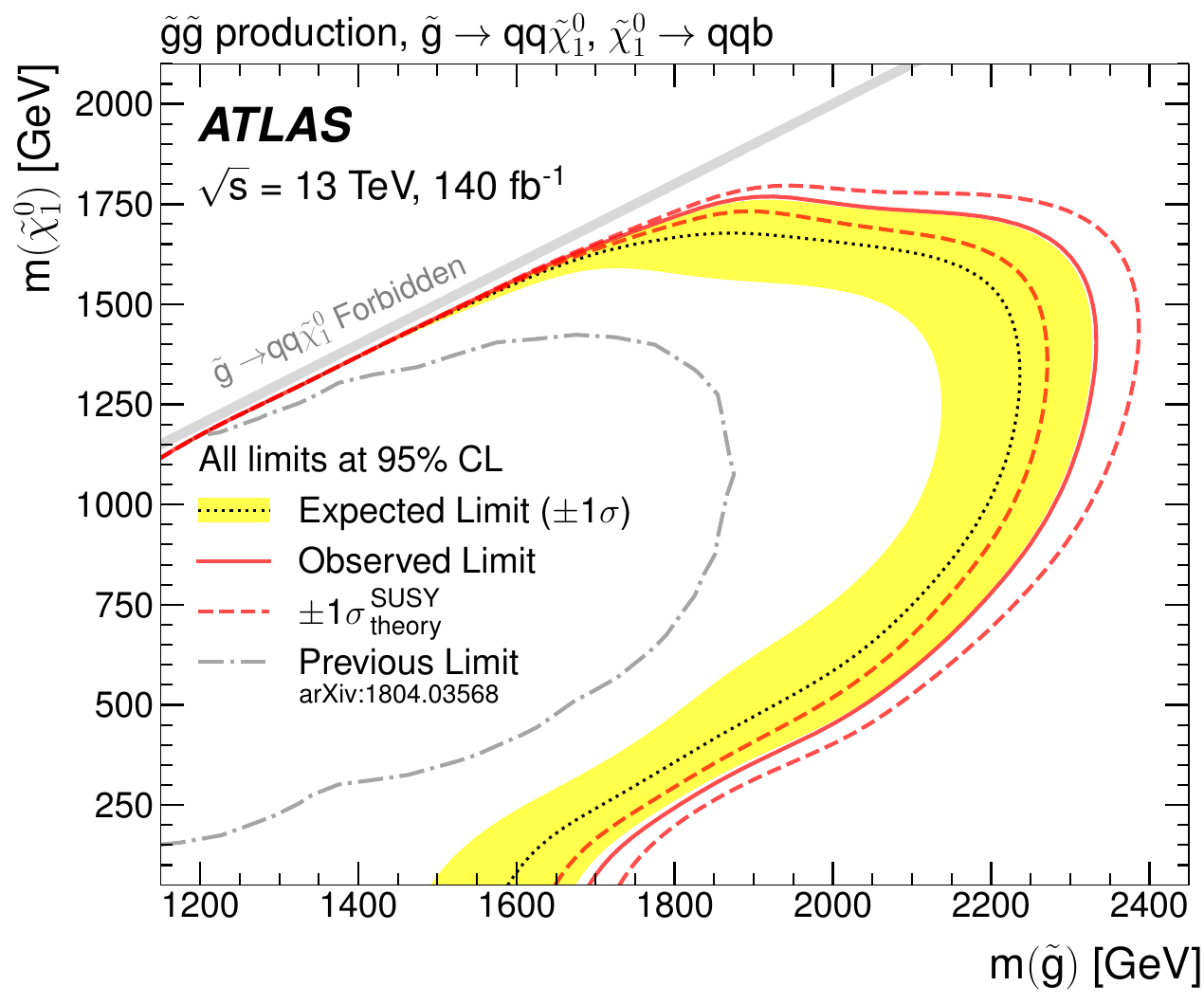} \label{fig:cca-obs-lim-2t5_b}}
 
\caption{
Observed and expected exclusion contours for the gluino cascade decay model with \subref{fig:cca-obs-lim-2t5_a} \UDS and \subref{fig:cca-obs-lim-2t5_b}  \UDB decays using the jet counting method.
The contours of the band around the expected limit are the $\pm 1 \sigma$ variations, including all uncertainties.
The dotted lines around the observed limit illustrate the change in the
observed limit as the nominal signal cross-section is scaled up and down by the theoretical uncertainty.
The diagonal line indicates the kinematic limit for the decay of the gluino.
}
\label{fig:cca-obs-lim-2t5}
\end{figure}

Exclusion limits as a function of the gluino mass for the gluino direct decay are shown in Figure~\ref{fig:cca-obs-lim-2t3_BestExpected}. Both the individual limits and the limit resulting from taking the best expected limit from each method are illustrated. Figure~\ref{fig:cca-obs-lim-2t5} shows the exclusion contours for the gluino cascade decay using the jet counting method.
For the jet counting method, the SR which provides the best expected sensitivity for a given gluino mass is used to set the limit.
Deriving the best expected limits from jet counting and mass resonance methods, gluinos with masses up to 1730 and 1800 \GeV are excluded in the gluino direct-decay models where the gluinos decay with 100\% BR into qqq (UDS coupling) and qqb (UDB coupling), respectively. 
For the gluino cascade decay model the limits are provided exclusively by the jet counting approach, where again the SR with the best expected sensitivity to a given signal mass scenario is used to set the limit.
Gluinos with masses up to 2230 (2340) \GeV are excluded for a neutralino with 1250 \GeV mass and UDS (UDB) coupling.
These results represent a significant improvement compared to previous analyses.
 
\pagebreak

% End of text imported from the .//sections/Results.tex input file

\FloatBarrier
 
\section{Conclusion}
% The next lines are included from the .//sections/Conclusions.tex input file
A search for R-parity-violating SUSY signals in events with multiple jets is performed with 140 $\text{fb}^{-1}$ of
proton–proton collision data at $\sqrt{s} = 13$~\TeV collected by the ATLAS detector at the LHC\@.
Two methods are used, a jet counting method searching for excess events in single-bin signal regions defined at high jet multiplicity and high $C$, and a mass resonance approach, which searches for a localised excess in the reconstructed gluino mass spectrum. A novel machine-learning approach is employed to address the combinatorial assignment problem and successfully reconstruct the gluino mass.
No significant excess is found in any signal region. Limits are set on the production of gluinos in the gluino direct
decay and cascade decay models in \UDD scenarios of RPV SUSY\@. In the gluino direct decay
model, gluinos with masses up to 1800~\GeV are excluded at 95\% CL\@. In the gluino cascade decay model, gluinos with masses as high as 2340~\GeV are excluded for a neutralino with 1250~\GeV mass.
Model-independent limits are also set on the visible cross-section times branching ratio in five overlapping signal regions.
These results improve upon the previously existing LHC limits owing to the larger luminosity, the introduction of event shape variables to suppress background, and the development of machine-learning techniques to assign jets to gluinos and reconstruct their mass.

% End of text imported from the .//sections/Conclusions.tex input file

\section*{Acknowledgements}

% The next lines are included from the .//acknowledgements/Acknowledgements.tex input file

We thank CERN for the very successful operation of the LHC and its injectors, as well as the support staff at
CERN and at our institutions worldwide without whom ATLAS could not be operated efficiently.
 
The crucial computing support from all WLCG partners is acknowledged gratefully, in particular from CERN, the ATLAS Tier-1 facilities at TRIUMF/SFU (Canada), NDGF (Denmark, Norway, Sweden), CC-IN2P3 (France), KIT/GridKA (Germany), INFN-CNAF (Italy), NL-T1 (Netherlands), PIC (Spain), RAL (UK) and BNL (USA), the Tier-2 facilities worldwide and large non-WLCG resource providers. Major contributors of computing resources are listed in Ref.~\cite{ATL-SOFT-PUB-2023-001}.
 
We gratefully acknowledge the support of ANPCyT, Argentina; YerPhI, Armenia; ARC, Australia; BMWFW and FWF, Austria; ANAS, Azerbaijan; CNPq and FAPESP, Brazil; NSERC, NRC and CFI, Canada; CERN; ANID, Chile; CAS, MOST and NSFC, China; Minciencias, Colombia; MEYS CR, Czech Republic; DNRF and DNSRC, Denmark; IN2P3-CNRS and CEA-DRF/IRFU, France; SRNSFG, Georgia; BMBF, HGF and MPG, Germany; GSRI, Greece; RGC and Hong Kong SAR, China; ISF and Benoziyo Center, Israel; INFN, Italy; MEXT and JSPS, Japan; CNRST, Morocco; NWO, Netherlands; RCN, Norway; MEiN, Poland; FCT, Portugal; MNE/IFA, Romania; MESTD, Serbia; MSSR, Slovakia; ARRS and MIZ\v{S}, Slovenia; DSI/NRF, South Africa; MICINN, Spain; SRC and Wallenberg Foundation, Sweden; SERI, SNSF and Cantons of Bern and Geneva, Switzerland; MOST, Taipei; TENMAK, T\"urkiye; STFC, United Kingdom; DOE and NSF, United States of America.
 
Individual groups and members have received support from BCKDF, CANARIE, CRC and DRAC, Canada; PRIMUS 21/SCI/017 and UNCE SCI/013, Czech Republic; COST, ERC, ERDF, Horizon 2020, ICSC-NextGenerationEU and Marie Sk{\l}odowska-Curie Actions, European Union; Investissements d'Avenir Labex, Investissements d'Avenir Idex and ANR, France; DFG and AvH Foundation, Germany; Herakleitos, Thales and Aristeia programmes co-financed by EU-ESF and the Greek NSRF, Greece; BSF-NSF and MINERVA, Israel; Norwegian Financial Mechanism 2014-2021, Norway; NCN and NAWA, Poland; La Caixa Banking Foundation, CERCA Programme Generalitat de Catalunya and PROMETEO and GenT Programmes Generalitat Valenciana, Spain; G\"{o}ran Gustafssons Stiftelse, Sweden; The Royal Society and Leverhulme Trust, United Kingdom.
 
In addition, individual members wish to acknowledge support from Chile: Agencia Nacional de Investigación y Desarrollo (FONDECYT 1190886, FONDECYT 1210400, FONDECYT 1230987); China: National Natural Science Foundation of China (NSFC - 12175119, NSFC 12275265); European Union: European Research Council (ERC - 948254), Horizon 2020 Framework Programme (MUCCA - CHIST-ERA-19-XAI-00), Italian Center for High Performance Computing, Big Data and Quantum Computing (ICSC, NextGenerationEU), Marie Sklodowska-Curie Actions (EU H2020 MSC IF GRANT NO 101033496); France: Agence Nationale de la Recherche (ANR-20-CE31-0013, ANR-21-CE31-0022), Investissements d'Avenir Idex (ANR-11-LABX-0012), Investissements d'Avenir Labex (ANR-11-LABX-0012); Germany: Baden-Württemberg Stiftung (BW Stiftung-Postdoc Eliteprogramme), Deutsche Forschungsgemeinschaft (DFG - CR 312/5-1); Italy: Istituto Nazionale di Fisica Nucleare (FELLINI G.A. n. 754496, ICSC, NextGenerationEU); Japan: Japan Society for the Promotion of Science (JSPS KAKENHI 22H01227, JSPS KAKENHI JP21H05085, JSPS KAKENHI JP22H04944); Netherlands: Netherlands Organisation for Scientific Research (NWO Veni 2020 - VI.Veni.202.179); Norway: Research Council of Norway (RCN-314472); Poland: Polish National Agency for Academic Exchange (PPN/PPO/2020/1/00002/U/00001), Polish National Science Centre (NCN 2021/42/E/ST2/00350, NCN OPUS nr 2022/47/B/ST2/03059, NCN UMO-2019/34/E/ST2/00393, UMO-2020/37/B/ST2/01043, UMO-2021/40/C/ST2/00187); Slovenia: Slovenian Research Agency (ARIS grant J1-3010); Spain: BBVA Foundation (LEO22-1-603), Generalitat Valenciana (Artemisa, FEDER, IDIFEDER/2018/048), La Caixa Banking Foundation (LCF/BQ/PI20/11760025), Ministry of Science and Innovation (RYC2019-028510-I, RYC2020-030254-I), PROMETEO and GenT Programmes Generalitat Valenciana (CIDEGENT/2019/023, CIDEGENT/2019/027); Sweden: Swedish Research Council (VR 2022-03845), Knut and Alice Wallenberg Foundation (KAW 2017.0100, KAW 2018.0157, KAW 2019.0447); Switzerland: Swiss National Science Foundation (SNSF - PCEFP2\_194658); United Kingdom: Leverhulme Trust (Leverhulme Trust RPG-2020-004); United States of America: Neubauer Family Foundation.

% End of text imported from the .//acknowledgements/Acknowledgements.tex input file

\clearpage

\printbibliography
 
\clearpage
% ATLAS Collaboration author list
% Reference date of SUSY-2019-24 is 2023-07-28
% Author list last updated on date 03-APR-24
% Data extracted on 03-Apr-2024 for paper reference SUSY-2019-24
% at 11:51am
 
\begin{flushleft}
\hypersetup{urlcolor=black}
{\Large The ATLAS Collaboration}

\bigskip

\AtlasOrcid[0000-0002-6665-4934]{G.~Aad}$^\textrm{\scriptsize 102}$,
\AtlasOrcid[0000-0002-5888-2734]{B.~Abbott}$^\textrm{\scriptsize 120}$,
\AtlasOrcid[0000-0002-1002-1652]{K.~Abeling}$^\textrm{\scriptsize 55}$,
\AtlasOrcid[0000-0001-5763-2760]{N.J.~Abicht}$^\textrm{\scriptsize 49}$,
\AtlasOrcid[0000-0002-8496-9294]{S.H.~Abidi}$^\textrm{\scriptsize 29}$,
\AtlasOrcid[0000-0002-9987-2292]{A.~Aboulhorma}$^\textrm{\scriptsize 35e}$,
\AtlasOrcid[0000-0001-5329-6640]{H.~Abramowicz}$^\textrm{\scriptsize 151}$,
\AtlasOrcid[0000-0002-1599-2896]{H.~Abreu}$^\textrm{\scriptsize 150}$,
\AtlasOrcid[0000-0003-0403-3697]{Y.~Abulaiti}$^\textrm{\scriptsize 117}$,
\AtlasOrcid[0000-0002-8588-9157]{B.S.~Acharya}$^\textrm{\scriptsize 69a,69b,m}$
% ...
\bigskip
\\

% ...
$^{29}$Physics Department, Brookhaven National Laboratory, Upton NY; United States of America.\\
% ...
$^{35}$$^{(a)}$Facult\'e des Sciences Ain Chock, R\'eseau Universitaire de Physique des Hautes Energies - Universit\'e Hassan II, Casablanca;$^{(b)}$Facult\'{e} des Sciences, Universit\'{e} Ibn-Tofail, K\'{e}nitra;$^{(c)}$Facult\'e des Sciences Semlalia, Universit\'e Cadi Ayyad, LPHEA-Marrakech;$^{(d)}$LPMR, Facult\'e des Sciences, Universit\'e Mohamed Premier, Oujda;$^{(e)}$Facult\'e des sciences, Universit\'e Mohammed V, Rabat;$^{(f)}$Institute of Applied Physics, Mohammed VI Polytechnic University, Ben Guerir; Morocco.\\
% ...
$^{49}$Fakult\"{a}t Physik , Technische Universit{\"a}t Dortmund, Dortmund; Germany.\\
% ...
$^{55}$II. Physikalisches Institut, Georg-August-Universit\"{a}t G\"ottingen, G\"ottingen; Germany.\\
% ...
$^{69}$$^{(a)}$INFN Gruppo Collegato di Udine, Sezione di Trieste, Udine;$^{(b)}$ICTP, Trieste;$^{(c)}$Dipartimento Politecnico di Ingegneria e Architettura, Universit\`a di Udine, Udine; Italy.\\
% ...
$^{102}$CPPM, Aix-Marseille Universit\'e, CNRS/IN2P3, Marseille; France.\\
% ...
$^{117}$Department of Physics, New York University, New York NY; United States of America.\\
% ...
$^{120}$Homer L. Dodge Department of Physics and Astronomy, University of Oklahoma, Norman OK; United States of America.\\
% ...
$^{150}$Department of Physics, Technion, Israel Institute of Technology, Haifa; Israel.\\
$^{151}$Raymond and Beverly Sackler School of Physics and Astronomy, Tel Aviv University, Tel Aviv; Israel.\\
% ...
$^{a}$ Also Affiliated with an institute covered by a cooperation agreement with CERN.\\
$^{b}$ Also at An-Najah National University, Nablus; Palestine.\\
$^{c}$ Also at Borough of Manhattan Community College, City University of New York, New York NY; United States of America.\\
$^{d}$ Also at Center for High Energy Physics, Peking University; China.\\
$^{e}$ Also at Center for Interdisciplinary Research and Innovation (CIRI-AUTH), Thessaloniki; Greece.\\
$^{f}$ Also at Centro Studi e Ricerche Enrico Fermi; Italy.\\
% ...
$^{m}$ Also at Department of Physics, King's College London, London; United Kingdom.\\
% ...

\end{flushleft}

% Created with Glance <Atlas.Glance@cern.ch>

\end{document}